%% file: 0paper.tex
\documentclass[format=sigconf, screen,nonacm]{acmart}
\input{00packages.tex}

\AtBeginDocument{%
  \providecommand\BibTeX{{%
    \normalfont B\kern-0.5em{\scshape i\kern-0.25em b}\kern-0.8em\TeX}}}

\begin{document}

%%
%% The "title" command has an optional parameter,
%% allowing the author to define a "short title" to be used in page headers.

\title[Does AI Coaching Prepare us for Workplace Negotiations?]{Does AI Coaching Prepare us for Workplace Negotiations?}

\author{Veda Duddu}
\orcid{0009-0001-6443-6239}
\affiliation{%
 \institution{University of Illinois Urbana-Champaign}
 \city{Urbana}
 \state{IL}
 \country{USA}}
\email{vduddu2@illinois.edu}

\author{Jash Rajesh Parekh}
\orcid{0000-0003-3310-4634}
\affiliation{%
 \institution{University of Illinois Urbana-Champaign}
 \city{Urbana}
 \state{IL}
 \country{USA}}
\email{jashrp2@illinois.edu}

\author{Andy Mao}
\orcid{0009-0007-6060-9730}
\affiliation{%
 \institution{University of Illinois Urbana-Champaign}
 \city{Urbana}
 \state{IL}
 \country{USA}}
\email{hanqim2@illinois.edu}

\author{Hanyi Min}
\orcid{0000-0002-0095-8513}
\affiliation{%
 \institution{University of Illinois Urbana-Champaign}
 \city{Urbana}
 \state{IL}
 \country{USA}}
 \email{hanyimin@illinois.edu}

\author{Ziang Xiao}
\orcid{0000-0003-3368-0180}
\affiliation{%
 \institution{Johns Hopkins University}
 \city{Baltimore}
 \state{MD}
 \country{USA}}
 \email{ziang.xiao@jhu.edu}

\author{Vedant Das Swain}
\orcid{0000-0001-6871-3523}
\affiliation{%
 \institution{New York University}
 \city{New York}
 \state{NY}
 \country{USA}}
 \email{v.dasswain@northeastern.edu}

\author{Koustuv Saha}
\orcid{0000-0002-8872-2934}
\affiliation{%
 \institution{University of Illinois Urbana-Champaign}
 \city{Urbana}
 \state{IL}
 \country{USA}}
\email{ksaha2@illinois.edu}

%%
%% By default, the full list of authors will be used in the page
%% headers. Often, this list is too long, and will overlap
%% other information printed in the page headers. This command allows
%% the author to define a more concise list
%% of authors' names for this purpose.
\renewcommand{\shortauthors}{}

%%
%% By default, the full list of authors will be used in the page
%% headers. Often, this list is too long, and will overlap
%% other information printed in the page headers. This command allows
%% the author to define a more concise list
%% of authors' names for this purpose.

%%
%% The abstract is a short summary of the work to be presented in the
%% article.
\input{0abstract}

%%
%% The code below is generated by the tool at http://dl.acm.org/ccs.cfm.
%% Please copy and paste the code instead of the example below.
%%


%%
%% Keywords. The author(s) should pick words that accurately describe
%% the work being presented. Separate the keywords with commas.

%%
%% This command processes the author and affiliation and title
%% information and builds the first part of the formatted document.
\maketitle

% \ifhidecomments
% \else
%     \thispagestyle{firststyle} % applies firststyle to first page
%     \pagestyle{allstyle}
% \fi

\input{1introduction.tex} %1.25 pages
\input{2relatedwork_new_v2}
\input{3study}
% \input{systemdesign}
% \input{3methods} %2 page
%3 page
% \input{semanticanalysis}
% \input{4findings}
\input{4findings_rq1}

\input{4findings_rq2}

\input{5discussion_v2}

\input{6limitations}
\input{7conclusion}

%%
%% The acknowledgments section is defined using the "acks" environment
%% (and NOT an unnumbered section). This ensures the proper
%% identification of the section in the article metadata, and the
%% consistent spelling of the heading.

% \begin{acks}

% \end{acks}

%%
%% The next two lines define the bibliography style to be used, and
%% the bibliography file.
\balance
\bibliographystyle{ACM-Reference-Format}
\bibliography{0paper}

% \includepdf[pages=-,pagecommand={},width=1.2\textwidth]{CSCW Job Satisfaction Reviewer Responses.pdf}

%%
%% If your work has an appendix, this is the place to put it.
% \appendix
% \input{7appendix.tex}

\end{document}

\endinput
%%
%% End of file `sample-acmsmall.tex'.

%% file: 00packages.tex
\usepackage{color, colortbl, xcolor}
\usepackage{url}
\usepackage{subcaption}
\usepackage{textcomp}
\usepackage{soul}
\usepackage{multirow}
\usepackage{enumitem}
\usepackage{mathtools}
\usepackage{siunitx}
\usepackage{array}
\usepackage{colortbl}
\usepackage{hhline}

\usepackage{booktabs} % For formal tables | Already in ACM
\usepackage{array}
\usepackage{xcolor}

\newcommand*{\rowstyle}[1]{% sets the style of the next row
  \gdef\@rowstyle{#1}%
  \@rowstyle\ignorespaces%
}

\newcolumntype{=}{% resets the row style
  >{\gdef\@rowstyle{}}%
}

\newcolumntype{+}{% adds the current row style to the next column
  >{\@rowstyle}%
}

\usepackage{arydshln}
\definecolor{LightGray}{gray}{0.97}
\definecolor{linkColor}{RGB}{6,125,233}
\definecolor{green}{rgb}{0.0, 0.65, 0.31}
\definecolor{bleudefrance}{rgb}{0.19, 0.55, 0.91}
\definecolor{ceruleanblue}{rgb}{0.16, 0.32, 0.75}
\definecolor{grey}{HTML}{969696}
\definecolor{violet}{HTML}{756bb1}
\definecolor{dgrey}{HTML}{01665e}
\definecolor{lgrey}{HTML}{5ab4ac}
\definecolor{dgreen}{HTML}{005a32}
\definecolor{purple}{HTML}{ae017e}

\definecolor{editCol}{HTML}{000000}
\definecolor{maskCol}{HTML}{c51b7d}
\definecolor{lrColor}{HTML}{8856a7}
\definecolor{trColor}{HTML}{d01c8b}
\definecolor{ctColor}{HTML}{4dac26}
\definecolor{brickred}{HTML}{f03b20}
\definecolor{improveCol}{HTML}{253494}
\definecolor{worsenCol}{HTML}{d7191c}
\definecolor{DarkBlue}{HTML}{00008B}
\definecolor{mscolor}{HTML}{01665e}
\definecolor{nmscolor}{HTML}{bf812d}
\definecolor{lgreen}{HTML}{ccece6}
\definecolor{dolive}{HTML}{308014}

\definecolor{editCol}{HTML}{000000}
\definecolor{maskCol}{HTML}{c51b7d}
\definecolor{lrColor}{HTML}{8856a7}
\definecolor{trColor}{HTML}{d01c8b}
\definecolor{ctColor}{HTML}{4dac26}
\definecolor{brickred}{HTML}{f03b20}
\definecolor{improveCol}{HTML}{253494}
\definecolor{worsenCol}{HTML}{d7191c}
\definecolor{lgreen}{HTML}{e0f3db}
\definecolor{dpink}{HTML}{CD1076}
\definecolor{pink}{HTML}{FED2D2}
\definecolor{soothinggreen}{HTML}{4dac26}
\definecolor{darkred}{HTML}{8B0000}

\definecolor{dblue}{HTML}{104E8B}
\definecolor{violet}{HTML}{8A2BE2}
\definecolor{mscolor}{HTML}{01665e}
\definecolor{nmscolor}{HTML}{d8b365}
\definecolor{deepgrey}{HTML}{525252}
\definecolor{dslate}{HTML}{2F4F4F}
\definecolor{dolive}{HTML}{556B2F}
\definecolor{teal}{HTML}{388E8E}
\definecolor{mscolor}{HTML}{01665e}
\definecolor{nmscolor}{HTML}{d8b365}

\definecolor{aicolor}{HTML}{018571}
\definecolor{occolor}{HTML}{ff7799}

\definecolor{srcolor}{HTML}{e34a33}
\definecolor{smcolor}{HTML}{253494}
\definecolor{srsmcolor}{HTML}{7fcdbb}
\definecolor{bothcolor}{HTML}{fe9929}
\definecolor{onecolor}{HTML}{018571}
\definecolor{marroon}{HTML}{881c1c}

\colorlet{tablerowcolor4}{gray!50} % Table row separator colour = 

\newcommand*{\textlabel}[2]{%
  \edef\@currentlabel{#1}% Set target label
  \phantomsection% Correct hyper reference link
  #1\label{#2}% Print and store label
}

\colorlet{tableheadcolor}{gray!25} % Table header colour = 25% gray
\colorlet{tablerowcolor}{gray!15} % Table row separator colour = 
\colorlet{tablerowcolor2}{gray!45} % Table row separator colour = 
\colorlet{tablerowcolor3}{gray!25} % Table row separator colour = 10% gray

\newcommand{\rowcolmedium}{\rowcolor{tablerowcolor2}}
\newcommand{\rowcollight}{\rowcolor{LightGray}} %

\newif{\ifhidecomments}
  \hidecommentsfalse 
\ifhidecomments
    \newcommand{\veda}[1]{}
    \newcommand{\andy}[1]{}
    \newcommand{\jash}[1]{}
    \newcommand{\haylee}[1]{}
    \newcommand{\ziang}[1]{}
    \newcommand{\vedant}[1]{}
    \newcommand{\koustuv}[1]{}
\else
    \newcommand{\veda}[1]{\textbf{\small\sffamily{\textcolor{DarkBlue}{[#1 -- Veda]}}}}
    \newcommand{\andy}[1]{\textbf{\small\sffamily{\textcolor{dgreen}{[#1 -- Andy]}}}}
    \newcommand{\jash}[1]{\textbf{\small\sffamily{\textcolor{dolive}{[#1 -- Jash]}}}}
    \newcommand{\haylee}[1]{\textbf{\small\sffamily{\textcolor{violet}{[#1 -- Haylee]}}}}
    \newcommand{\ziang}[1]{\textbf{\small\sffamily{\textcolor{brickred}{[#1 -- Ziang]}}}}
    \newcommand{\vedant}[1]{\textbf{\small\sffamily{\textcolor{soothinggreen}{[#1 -- Vedant]}}}}
    \newcommand{\koustuv}[1]{\textbf{\small\sffamily{\textcolor{dpink}{[#1 -- Koustuv]}}}}
  \fi

\newcommand{\trc}{\texttt{Trucey}}
\newcommand{\cgpt}{\texttt{ChatGPT}}
\newcommand{\hbk}{\texttt{Handbook}}

\renewcommand{\textrightarrow}{$\rightarrow$}

\newcommand{\n}[1]{$\mathtt{#1}$}

\colorlet{tableheadcolor}{gray!25} % Table header colour = 25% gray
\colorlet{tablerowcolor}{gray!5} % Table row separator colour = 10% gray

\definecolor{neutralCol}{HTML}{dd1c77}
\definecolor{neutralGreen}{HTML}{31a354}
\definecolor{NewBlue}{HTML}{1879ba}
\definecolor{bleudefrance}{rgb}{0.19, 0.55, 0.91}  
\definecolor{AfTrColor}{HTML}{0868ac}  
\definecolor{BfTrColor}{HTML}{a8ddb5}  

\definecolor{AfCtColor}{HTML}{b10026}  
\definecolor{BfCtColor}{HTML}{fd8d3c}

\graphicspath{ {figures/} }

\newcommand{\para}[1]{\vspace{0.5em}\noindent\textbf{#1}~}

%% file: 0abstract.tex
% \newcommand\jp[1]{\textcolor{blue}{jash: #1}}

\begin{abstract}
%how do we change the anonymous authors?

% david: check word count
% confusion about domain-specific terms, such as "fear" 
% need a bit more context before introducing Trucey. can be less specific in the statistical findings, e.g. no need to specify d=0.08. just say significance, marginal significance, w/e
% i like how the ending wraps up the study by contexutalizing assumptions 

Workplace negotiations are undermined by psychological barriers, which can even derail well-prepared tactics. AI offers personalized and always-available negotiation coaching, yet its effectiveness for negotiation preparedness remains unclear. We built \trc{}, a prototype AI coach grounded in Brett’s negotiation model. We conducted a between-subjects experiment (N=267), comparing \trc{}, \cgpt{}, and a traditional negotiation \hbk{}, followed by in-depth interviews (N=15). 
While \trc{} showed the strongest reductions in fear relative to both comparison conditions, the \hbk{} outperformed both AIs in usability and psychological empowerment. Interviews revealed that the \hbk{}’s comprehensive, reviewable content was crucial for participants’ confidence and preparedness. In contrast, although participants valued AI's rehearsal capability, its guidance often felt verbose and fragmented—delivered in bits and pieces that required additional effort—leaving them uncertain or overwhelmed. 
These findings challenge assumptions of AI superiority and motivate hybrid designs that integrate structured, theory-driven content with targeted rehearsal, clear boundaries, and adaptive scaffolds to address psychological barriers and support negotiation preparedness.

\end{abstract}

%% file: 1introduction.tex
\section{Introduction} \label{section:intro}
Workplace negotiations---such as requesting promotions, asserting boundaries, or challenging authority---are recognized to be difficult and emotionally fraught~\cite{Bradley_Campbell_2016,grandey2000emotional}. 
Existing theory and practice have largely emphasized on negotiation strategies and tactics, focusing on structural, processual, and communicative components~\cite{Baber_2022, nadler2003learning}. However, upward negotiation is especially complex due to hierarchical power imbalances, institutional norms that reinforce authority structures, and workers' concerns about retaliatory consequences~\cite{morrison2000organizational, Cortina_Magley_2003}. 
Such concerns often extend to fears of being negatively perceived within the workplace, damaging relationships, or jeopardizing career prospects~\cite{morrison2000organizational, Milliken_Morrison_Hewlin_2003}. 
These fears create psychological barriers---increasing anxiety and reducing confidence---that undermine negotiation outcomes regardless of strategy, often leading to lowered expectations and premature withdrawal from bargaining~\cite{Brooks_Schweitzer_2011,sullivan2006negotiator}.
% , such as increased anxiety and diminished confidence, undermining negotiation outcomes regardless of strategy, frequently resulting in lower expectations and premature withdrawal in bargaining~\cite{Brooks_Schweitzer_2011,sullivan2006negotiator}.
% Workers are known to worry about their perception within the firm, damaged relationships, career damage, and more~\cite{morrison2000organizational, Milliken_Morrison_Hewlin_2003}. 

% within the workplace---factors which are known to affect negotiation outcomes, inrrespective of negotiation strategy, often leading to worse outcomes because of workers' low expectations and earlier exit in bargaining~\cite{Brooks_Schweitzer_2011,sullivan2006negotiator}. 
% Research demonstrates how anxiety affects the outcomes of negotiators irrespective of their choice of strategy, often leading to worse outcomes due to their low expectations and exiting earlier in bargaining~\cite{Brooks_Schweitzer_2011}. 
Furthermore, these psychological barriers can undermine self-efficacy and self-esteem, which in turn may not only lead to \textit{negotiation impasse}, as well as but also affect a worker's wellbeing, satisfaction, and performance~\cite{o2001distributive}.
% self-efficacy and confidence have been established as consequences of the very same psychological barriers, affecting future negotiations~\cite{Connor_Arnold_2001}. 
% \veda{might remove/find a better citation for the previous line}
Addressing such psychological barriers is therefore essential, as fear and low confidence often prevent workers from effectively enacting their negotiation strategies, making this challenge as critical as strategizing negotiation tactics. %the consideration as important as strategizing negotiation tactics. 
% Therefore, addressing psychological barriers is as critical as tactical knowledge, as fear and low confidence prevent employees from effectively implementing the negotiation strategies they possess.

% Traditional negotiation coaching, either in the form of self-guided videos and resources, or as in-person coaching, has been known to improve outcomes such as promotions and raises~\cite{Baber_2022}.
Traditional negotiation coaching---whether through self-guided videos and resources or in-person training---has been shown to improve critical workplace outcomes such as promotions and salary increases~\cite{Baber_2022,brett2016negotiation}.
% \ziang{When we are referring to traditional coaching, do we mean in-person coaching or also self-guided videos or books?} 
Yet, these forms of coaching are typically static, one-size-fits-all, and available only in scheduled sessions, limiting access round-the-clock and personalized support that workers may need. 
Further, these approaches primarily emphasize strategies of negotiation and communication, without directly coaching individuals through their psychological barriers. 
AI offers an opportunity to extend these supports---recent advances in generative AI (genAI), especially large language models, have enabled seemingly personalized and natural language communication, and early work highlights the promise of AI-assisted rehearsal and emotion regulation for workers~\cite{dasswain2025ai,shaikh2024rehearsal,das2024teacher,mckendrick2023virtual}.  
% However, work such as Erlei et al \cite{erlei2022s} also showcases how humans avoid AI in bargaining despite economic benefits, making rehearsal with the help of AI a complex decision. 
% \koustuv{here we may also want to cite Ujjwal's works on bargaining}
That said, it remains an open question: \textbf{How effective AI can be for negotiation coaching and what new challenges it may introduce}. 

% be effective, allowing workers to achieve promotions and raises~\cite{Baber_2022}. However, these traditional systems are not available to provide round-the-clock support. Therefore, workers may not have support when they require it the most. AI offers an opportunity to provide more on-demand training and support in various workplace situations~\cite{dasswain2025ai,shaikh2024rehearsal,das2024teacher}. 

Building on the momentum of AI-assisted workplace technologies~\cite{park2022designing,roemmich2023emotion,kawakami2023wellbeing, kaur2022didn}, early investigations into AI support for interpersonal conflicts and related domains have highlighted both the
% AI-assisted negotiations have revealed encouraging results, 
% particularly the 
strengths and gaps in tactical skill development and knowledge acquisition~\cite{falcao2024making,ma2025chatgpt,shea2024ace}. ~\citeauthor{shaikh2024rehearsal} emphasized conflict resolution strategies, and~\citeauthor{ma2025chatgpt} emphasized the need to incorporate expert knowledge.  
% Yet, despite these advances in theory and practice, prior investigations focus primarily on skills rather than psychological aspects.
% However, psychological factors such as self-efficacy are key in negotiation. 
Yet despite these advances, prior work has focused primarily on skills rather than psychological factors, even though psychological aspects such as self-efficacy are key in negotiation.
Organizational psychology research notes that negotiators with greater self-efficacy are more resilient to setbacks, sustain a willingness to negotiate, and maintain a more positive orientation toward the negotiation process~\cite{o2001distributive}. 
This highlights the critical need for AI systems that not only strengthen tactical preparation, but also directly reduce psychological barriers and empower individuals throughout their careers.
Therefore, motivated by the limitations of traditional and AI-based negotiation training, our work is guided by the following research questions (RQs):

 \begin{enumerate}
     % \item[\textbf{RQ1}:] How does theory-driven AI negotiation preparation affect psychological readiness for workplace negotiations?
     \item[\textbf{RQ1}:] How does a theory-driven AI negotiation coach influence psychological readiness for workplace negotiations?
     % compared to generic AI approaches? \veda{should it be does/ how does?}
     % \item[\textbf{RQ2}:] What psychological mechanisms explain the differential effects of static versus dynamic negotiation preparations?
     % \item[\textbf{RQ2}:] What psychological mechanisms explain the impact of a theory-driven AI negotiation coach for workplace negotiations?
     \item[\textbf{RQ2}:] How do individuals perceive and make sense of a theory-driven AI negotiation coach?
     % \item[\textbf{RQ2}:] Are there psychological mechanisms that explain the differential effects of static vs dynamic negotiation preparation interactions?
\end{enumerate}

% These RQs enable us to examine both the efficacy of AI coaching for workplace negotiations and the underlying psychological mechanisms that shape its impact. 
For our study, we designed \textbf{\trc{}}, an AI-powered interactive prototype that aims to prepare individuals for complex workplace negotiations by targeting both tactical and psychological barriers. 
% \ziang{Do we want to talk about the interaction with Trucey, e.g., how the coaching is delivered? It may help audience to understand the difference between conditions.} 
Grounded in human-centered AI and Industrial-Organizational (I/O) psychology---particularly~\citeauthor{brett2016negotiation}'s negotiation framework---\trc{} was built using few-shot learning on fine-tuned GPT-4.1.
\trc{} delivers negotiation coaching through a structured conversational interaction: for a negotiation scenario, users receive tailored prompts to articulate goals and challenges, and then rehearse role-based exchanges with \trc{} acting as a counterpart. Throughout, the AI provides feedback that highlights potential strategies, surfaces power dynamics, and invites reflection on confidence and emotional responses.
% we employed few-shot learning on fine-tuned GPT-4.1.
% , to incorporate four key features
Such an interaction was enabled by incorporating four key design features---1) \textit{situational calibration}, 2) \textit{role-based simulation}, 3) \textit{dynamic contextual layering}, and 4) \textit{iterative response alignment}. 
% \ziang{I think we do need to explain what Trucey is doing; otherwise, those features are not clear. (you can also consider changing the name). } 
These design features help create negotiation scenarios that embed contextual and power dynamics known to trigger psychological barriers, providing a rigorous environment to examine how AI coaching interventions can influence negotiation readiness.

% \ziang{should we mention human-centered AI design here? It implies formative study and iterative design.} 
% Industrial-Organizational (I/O) psychology and human-centered AI design. 
% \trc{} aims at preparing individuals for complex workplace negotiations by targeting both tactical and psychological barriers.
% To unpack further, the design of \trc{} is guided by 
% \trc{}'s design is guided by \citeauthor{brett2016negotiation}'s negotiation framework, which emphasizes that outcomes are shaped less by personality traits and more by interaction patterns, power asymmetries, and strategy alignment~\cite{brett2016negotiation}. 
% Building on this foundation, we 
% Building on \citeauthor{brett2016negotiation}'s framework, 
% \textit{situational calibration} to account for context and power dynamics, 2) \textit{role-based simulation} to practice tone and stance, 3) \textit{dynamic contextual layering} to scaffold information in manageable cognitive loads, and 4) \textit{iterative response alignment} to adapt coaching based on user feedback.
% These features create 
% \koustuv{All of these content in the Intro might make it too long---we can move some of these to the methods where we describe and motivat Trucey}

% \koustuv{We conducted a pre-registered experimental study}

To address our RQs, we first conducted a pre-registered experimental study with $N$=267 participants across three conditions---1) \textbf{\trc{}}, an AI-powered coach integrating negotiation theory ($N$=134); 2) \textbf{\cgpt{}}, a generic AI without theory-driven negotiation scaffolding ($N$=66); and 3) \textbf{\hbk{}}, a static but theory-based negotiation resource without AI interactivity ($N$=67). 
Participants completed pre- and post-task surveys measuring psychological empowerment, self-efficacy, and preparedness of negotiation. 
This design enabled us to disentangle the contributions of AI interactivity and theory-driven negotiation support, providing a rigorous test of how these elements shape psychological readiness for workplace negotiation.
In addition, we conducted an in-depth interview study ($N$=15) to explore participants' mental models, their perceived strengths and limitations of each condition, and how AI- or theory-driven support shaped their confidence and approach to workplace negotiation. 
This qualitative component complemented the experimental findings by providing richer insights into the experiential and interpretive dimensions of AI-assisted negotiation preparation.

Our experimental study demonstrated that participants significantly favored the theory-based \hbk{} over both AI coaching tools on usability and psychological empowerment, but \trc{} significantly reduced fear compared to generic AI coaching (Cohen's $d$=-0.23). 
We conducted linguistic analyses to examine how the language of interactions was associated with the psychological outcomes. 
Interviews revealed that participants preferred the \hbk{}'s comprehensive content because it provided greater control over their preparation process and allowed them to develop personalized strategies at their own pace. In contrast, they found AI coaching fragmented and insufficient for preparing across scenarios, though \trc{}'s rehearsal component helped them practice uncomfortable conversations.

Taken together, our work makes both empirical and conceptual contributions.
First, we develop and evaluate \trc{}, a theory-driven AI negotiation coach that targets psychological factors such as self-efficacy, fear reduction, and empowerment in negotiation preparation.
Second, we provide empirical evidence that the value of AI in negotiation is situated—shaped by stress, authenticity concerns, and contextual constraints—rather than universally beneficial. Our findings challenge common assumptions about the superiority of personalized, conversational AI, showing that in high-stakes workplace contexts, user control and access to comprehensive resources may be more critical for psychological empowerment than conversational guidance alone.
Finally, this work bears methodological and design implications for AI-mediated communication, extending the scope of workplace technologies by emphasizing strategies that balance efficiency with authenticity, autonomy, and inclusivity to better support confidence, resilience, and equitable participation in professional interactions.

%% file: 2relatedwork_new_v2.tex
\section{Related Work}

\subsection{Workplace Negotiation and Psychological Barriers}

Workplace negotiations—such as requesting promotions, asserting boundaries, or challenging authority—are shaped by both structural and psychological barriers that make self-advocacy difficult. Employees often withhold input due to fears of retaliation, damaging relationships, or jeopardizing career prospects~\cite{morrison2000organizational,detert2007leadership}, concerns that are amplified by hierarchical power asymmetries~\cite{keltner2003power,galinsky2006power}. Avoidance is further reinforced by conflict dynamics, as both relationship and task conflict reliably undermine performance~\cite{de2003task}, and by social norms such as gendered backlash for self-advocacy~\cite{bowles2007social} and cultural expectations of harmony~\cite{ting1988face}.

% Individual psychological traits critically shape negotiation engagement and outcomes. 
Self-efficacy predicts willingness to initiate and persistence through challenges~\cite{bandura1997self}, whereas anxiety and low confidence reduce joint value creation~\cite{brooks2014get}. Socio-emotional skills, including emotion recognition and emotional intelligence, positively influence negotiation performance~\cite{elfenbein2002emotional,sharma2013development}. Importantly, negotiation success extends beyond economic outcomes: subjective experiences regarding process, relationship, and self-evaluation independently predict long-term negotiation behavior~\cite{curhan2006people,curhan2010objective}. Incremental beliefs about ability further enable negotiators to achieve better outcomes than those with fixed mindsets~\cite{kray2007implicit}.

These insights underscore the importance of theory-driven frameworks to understand and intervene in negotiation behavior. Brett’s synthesis of distributive and integrative approaches highlights how information sharing, culture, and cognitive biases systematically influence outcomes~\cite{brett2016negotiation}. Complementing this, \citeauthor{Bradley_Campbell_2016}'s work on conversational staging shows how negotiation unfolds across temporal phases, with different barriers and opportunities emerging at each stage~\cite{Bradley_Campbell_2016}. Together, these frameworks provide a structured account of negotiation dynamics that can guide both research and the design of interventions aimed at mitigating psychological barriers.

\subsection{Human–AI Interaction: Strengths, Weaknesses, and Challenges}

AI systems hold promise for enhancing human capabilities, productivity, and collaboration
extending human capacities via automation, personalized feedback, and decision support~\cite{das2024teacher,shaikh2024rehearsal}. Yet many AI systems produce errors or outputs misaligned with organizational goals~\cite{raji2022fallacy}, and anticipating unintended consequences remains challenging~\cite{boyarskaya2020overcoming,coston2022validity,raji2020closing, varanasi2023currently}. 
Recent research underscores that achieving human–AI alignment requires more than technical fixes: systems must adapt to users' goals, practices, and values, and misalignments can erode trust, limit agency, or distort decision-making~\cite{cai2019human,inkpen2023advancing,wang2021mutual, kim2024beyond}. 
Scholars have responded with taxonomies of failures, design guidelines, and frameworks for transparency and accountability~\cite{amershi2019guidelines,chancellor2019taxonomy,raji2020closing,liao2020questioning,arrieta2020explainable,mitchell2019model,gebru2021datasheets,sokol2020explainability,shen2024towards}, emphasizing that reliable, ethical, and usable AI requires more than algorithmic performance alone.

Human–AI interactions also introduce distinct cognitive and social dynamics. Users often calibrate trust differently with AI, sometimes resisting algorithmic advice even when it is superior~\cite{dietvorst2015algorithm,glikson2020human, wang2019human, gero2020mental, sharma2023investigating}. People may sacrifice financial gain to avoid bargaining with AI, preferring human negotiation partners~\cite{erlei2022s,erlei2020impact}, and poorly calibrated trust can reduce effectiveness or create confusion in high-stakes or socially complex contexts~\cite{okamura2020adaptive,hemmer2021human}. While traditional explanation methods partially improve understanding, few reliably enhance appropriate reliance, prompting exploration of alternatives like analogy-based explanations leveraging commonsense knowledge~\cite{he2022walking, szymanski2025limitations, gebreegziabher2025supporting, sharma2023cognitive}. Personalized AI can strongly influence judgment over time~\cite{jahanbakhsh2023exploring}, and LLM-assisted systems show that trust in competence, integrity, and benevolence can reshape experiences, sometimes masking newly introduced burdens~\cite{jo2025ai}.

These challenges are amplified in organizational settings, where AI can affect power asymmetries, wellbeing, and professional relationships. Decision-support AI may influence promotions, evaluations, or workload distribution, with cascading effects on behavior and perceptions~\cite{lee2015working,ajunwa2017limitless}. Understanding these dynamics is essential for designing AI to support negotiation or other workplace interactions without exacerbating psychological barriers.

Taken together, while AI holds potential to augment human capabilities, realizing these benefits requires careful attention to trust, interpretability, human agency, and socio-organizational context. These insights motivate the development of AI interventions that not only provide tactical support but also actively scaffold psychological readiness and decision-making in complex workplace scenarios. In this work, we examine human-AI interaction in workplace negotiations, focusing on how users perceive AI's role as a negotiation coach, alignment with their needs and agency, and how these perceptions shape the effectiveness of AI as both tactical support and psychological scaffold. 

\subsection{Human-centered Technologies for Workplace}

A rich body of HCI and CSCW research explores technologies designed to support professional communication, skill development, and wellbeing in organizational settings~\cite{epstein2016reconsidering,nepal2025survey, saha2019libra,dasswain2020social,dasswain2020culture}. 
These human-centered technologies increasingly integrate sensing, collaborative, and AI-driven approaches to address complex workplace needs. 
Workplace AI applications demonstrate both significant promise and inherent limitations~\cite{park2022designing,kawakami2023wellbeing,roemmich2023emotion,kaur2022didn,chowdhary2023can}. 
On the one hand, AI can streamline routine tasks such as email management and meeting scheduling~\cite{mark2016email,howe2022design}, and a range of tools and methods~\cite{shea2024ace, wilhelm2025managers, khadpe2024discern, kuo2023understanding} highlight potential to scaffold workplace conversations~\cite{jahanbakhsh2017you}.

On the other hand, current systems remain constrained: they often lack the contextual awareness and interpersonal sensitivity required for organizational life, leading to mismatches between what AI can deliver and the nuanced demands of workplace communication and collaboration~\cite{ackermann2011strategic,roemmich2023emotion,das2023algorithmic}.
Related research on AI-mediated communication highlights opportunities to reduce workload and stress while also raising concerns about fairness, privacy, and inclusion~\cite{kawakami2023wellbeing,das2023algorithmic,kaur2022didn}. Within these systems, organizational power dynamics and competing stakeholder interests further complicate adoption and use~\cite{lee2015working,ajunwa2017limitless,chowdhary2023can,gomez2020taxonomy, lallemand2024trinity}, shaping whether interventions are perceived as empowering or threatening.
% For example, AI can improve productivity in areas such as email management and meeting scheduling~\cite{mark2016email,howe2022design}, and tools like ACE~\cite{shea2024ace} and CommCoach~\cite{wilhelm2025managers} revealed the potential to support workplace conversations.
% However, current AI systems often fall short of meeting the nuanced demands of organizational life, especially for tasks that require contextual awareness and interpersonal understanding~\cite{ackermann2011strategic}.
% For example, AI can enhance productivity in domains like email management and meeting scheduling~\cite{mark2016email,howe2022design}, and systems such as ACE~\cite{shea2024ace} and CommCoach~\cite{wilhelm2025managers} show potential for supporting workplace conversations. 
% Yet, research consistently reveals substantial gaps between current AI capabilities and the nuanced requirements of organizational life, particularly for tasks demanding contextual sensitivity and interpersonal understanding~\cite{ackermann2011strategic}.
% Prior work has developed human-centered technologies to support worker wellbeing, such as wearables and ambient systems that detect stress, monitor workload, or provide feedback on work–life balance~\cite{epstein2016reconsidering,nepal2025survey, saha2019libra}. 

These tensions help explain why prior work revealed professionals' preference for human advice networks over algorithmic recommendations~\cite{longoni2019resistance}, and why concerns about over-reliance surface when automation is introduced into socially complex tasks such as negotiation~\cite{carr2014glass}. The adoption of AI in the workplace raises ethical, social, and practical challenges at both individual and organizational levels~\cite{wagner2021measuring,boyd2012critical}. 
Organizational culture plays a critical mediating role in whether such systems are adopted successfully, and ethical issues demand accountability frameworks to protect sensitive interpersonal data~\cite{crawford2014big,jobin2019global}.

Our work extends this body of research by focusing on AI as a preparation coach for workplace negotiations rather than a direct participant. This framing preserves user agency while addressing key psychological barriers---such as fear and self-efficacy---that hinder negotiation preparedness. By empirically examining how AI-mediated coaching shapes psychological readiness, we contribute evidence that complements existing workplace technologies while highlighting design principles that emphasize autonomy, inclusivity, and resilience. More broadly, our findings add to ongoing CSCW and HCI discourse on how human-centered technologies can support socially complex, high-stakes interactions.

%% file: 3study.tex
\section{Study Design and Methods}

% \koustuv{Let's have a rigorous cleanup of this section. A reader will be lost with the back-and-forth of information in here. I am providing a skeleton structure, could you start filling out the information into these?}

% \subsection{Overview of Study Design}

\begin{figure*}[t]
    \centering
    \includegraphics[trim={0cm 5cm 0cm 0cm}, clip, width=1.95\columnwidth]{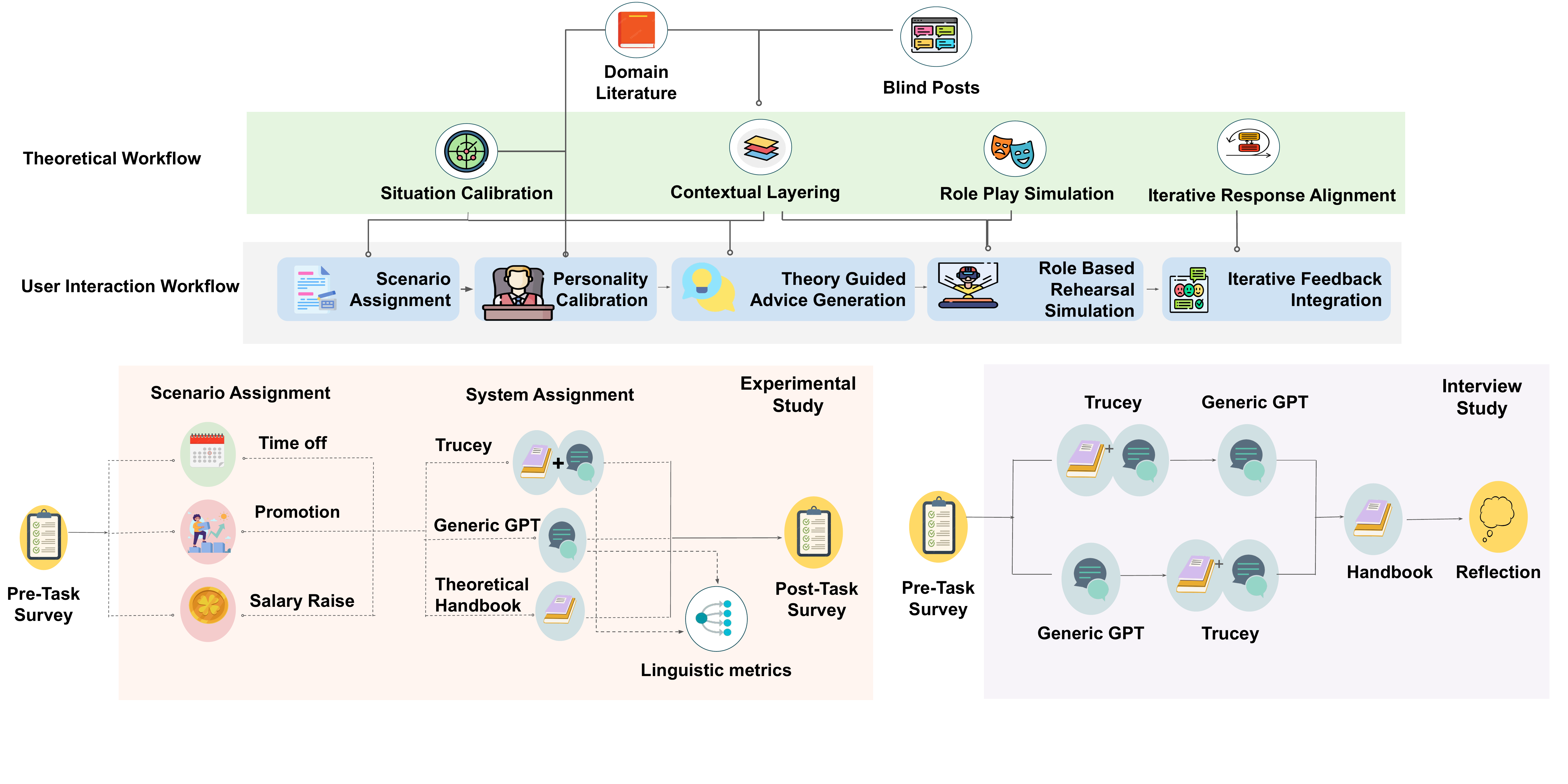}
    \Description[table]{This diagram provides a schematic overview of the study's design and the underlying architecture of Trucey, the AI-powered negotiation coach prototype. The study employed a mixed-methods approach, integrating a large-scale quantitative experiment with a smaller-scale qualitative interview study to gain comprehensive insights into the effectiveness and user perceptions of AI-based negotiation support. Theoretical Workflow and User Interaction Workflow. The system's design is grounded in negotiation theories from organizational science, particularly the frameworks of Brett and Thompson and Bradley and Campbell. Trucey operationalizes these theories through a four-mechanism architecture: Situation Calibration, Contextual Layering, Role Play Simulation, and Iterative Response Alignment. The user interacts with Trucey through a structured five-stage workflow that includes Scenario Assignment, Personality Calibration, Theory-Guided Advice Generation, Role-Based Rehearsal Simulation, and Iterative Feedback Integration.Experimental Study and Interview Study. For the experimental study, participants were randomly assigned to three conditions: Trucey, a generic GPT chatbot and theoretical handbook. They completed pre- and post-task surveys to measure psychological metrics, and the language of their interactions with the AIs was also analyzed. The interview study, conducted with a separate group of participants, involved a counterbalanced experience with the two AI conditions and a review of the Handbook, followed by in-depth semi-structured interviews to explore their perceptions and experiences
}
    \caption{A schematic overview of \trc{} and our study design.}
    \label{fig:schematic}
\end{figure*}

We designed a preregistered mixed-methods study.
% \footnote{\hl{[Insert OSF Link]}}, 
consisting of a quantitative three-condition between-subject experiment $N$=267 followed by a qualitative semi-structured interview with $N$=15 participants to understand mechanisms and gain deeper insights (\autoref{fig:schematic} provides a schematic overview of our study). 
% Our study was approved by the Institutional Review Board (IRB) at our university.
In this section, we describe our study design, participant recruitment, and methodology.

\subsection{Designing an Interactive Prototype: \trc{}}

% \koustuv{this should be a few paragraph subsection on what you designed, how you designed, using Brett's theory. We do not need to talk about anything related to experimental study here.}

\subsubsection{Theoretical Foundation}
Our study centered on the design of an AI-powered interactive prototype for workplace negotiation, \trc{}. 
\trc{}'s design was guided by two complementary theoretical frameworks from the organizational science literature. 
First,~\citeauthor{brett2016negotiation}'s negotiation framework emphasizes how contextual factors and power asymmetries shape the effectiveness of negotiation strategies~\cite{brett2016negotiation}. \citeauthor{brett2016negotiation}'s focus on situational dynamics, rather than fixed personality traits, aligned with our objective of supporting users in managing aspects of negotiation that are within their control. 
Second, we drew inspiration from \citeauthor{Bradley_Campbell_2016}'s framework for managing difficult conversations~\cite{Bradley_Campbell_2016}.
\citeauthor{Bradley_Campbell_2016} highlights three critical phases—preparation, in-conversation management, and post-interaction reflection \cite{Bradley_Campbell_2016}. This structured approach provides a process-oriented lens, ensuring that negotiation support extends beyond isolated tactics to scaffold the entire interaction.

Together, these two frameworks informed the design of \trc{} as a theory-driven AI coach that integrates strategic guidance with conversational process support, enabling users to reflect, adapt, and navigate negotiations more effectively.

% For our study, we designed an AI-powered interactive prototype for workplace negotiation, \trc{}. 
% \trc{} was designed as a theory-driven AI coaching system, grounded in established frameworks from negotiation research, particularly \citeauthor{brett2016negotiation}'s negotiation framework.
% ~\cite{brett2016negotiation}. The theoretical foundation for this study was Brett \& Thompson's negotiation framework, 
% chosen for its empirically grounded insights into how contextual factors and power asymmetries shape negotiation strategy effectiveness \cite{brett2016negotiation}. Crucially, their framework's emphasis on situational factors, rather than stable personality traits, is congruent with the study's objective of designing an AI system that enables users to navigate controllable contextual elements without attempting to alter inherent individual characteristics.

% The design also drew inspiration from Bradley and Campbell's framework for managing difficult conversation \cite{Bradley_Campbell_2016}. The framework emphasized three critical phases - preparation, in-conversation management and post-interaction reflection. The structured approached enables \trc{} to support users throughout the entire negotiation process rather than isolated advice. The combination of these frameworks enabled the creation of an AI system that aims to address both the strategic aspects and the conversational dynamics at play. 

\subsubsection{Prototype Architecture}
\trc{} translates the above theoretical foundations into practice through four key mechanisms that work together to provide contextually appropriate coaching, as listed below:

\para{Situational Calibration} ensures that shared contextual understanding enhances coaching efficacy in high-stakes scenarios, thereby mitigating the provision of generic advice incongruent with the user's specific circumstances. 

\para{Role-based Simulation} ensures the AI adopts an appropriate perspective during practice, thereby facilitating a realistic conversational dynamic for the user.

\para{Contextual Layering} emulates the incremental information delivery characteristic of effective human coaching across multiple dialogue turns. This approach reduces cognitive load and aims to prevent user overwhelm.

\para{Iterative Response Alignment} integrates active listening behaviors to maintain shared understanding and facilitate a situationally responsive coaching system.

% \begin{figure*}[t]
%     \centering
%     % \includegraphics[width=2\columnwidth]{figures/framework-diagram-v2.png}
%     % \includegraphics[width=\columnwidth]{figures/schematic_figure_v4.pdf}
%     % \koustuv{Add the figure here!}
%     % \caption{Overview of our study design for identifying and mitigating social stereotypes in T2I output.}
%         \label{fig:study_design}
% \end{figure*}

% The mechanisms integrates through a unified prompting architecture that adapts \citeauthor{brett2016negotiation}'s negotiation strategies to individual user contexts. The system processes contextual variables including negotiation topic, relationship quality metrics, supervisor's personality profile and prior discussion history to generate theoretically grounded but personalized coaching responses. 

\subsubsection{Interaction Design}
\trc{}'s interaction design followed a structured five-stage workflow, combining \textit{advice} and \textit{rehearsal}, supporting both the preparation and practice components essential for effective negotiation coaching. 
% five-phase workflow, as listed below:

% \begin{enumerate}
    \para{Scenario Assignment.} First, \trc{} would presents users with workplace negotiation scenarios with specific contextual details to make the situation more realistic. For example, scenarios contain details such as:
    
    \begin{quote}
    \small
    \sffamily
    \textbf{Problem to discuss:} Asking For Time Off
    
    \textbf{Person you are talking to:} Your boss
    
    \textbf{Your relationship:} You have had a bad 
    relationship for two months
    
    \textbf{Previous discussions:} You have discussed this before
    Work context: You are paid  hourly wages
    \end{quote}
    
    %containing pre-structured contextual variables including relationship duration and quality metrics.\koustuv{what does this mean? elaborate/give example.}
    
    \para{Personality Calibration.}  Second, \trc{} asked users to complete a structured assessment of their supervisor's behavioral patterns using Big-Five personality traits~\cite{soto2017next}, which research shows significantly impact negotiation outcomes~\cite{brett2016negotiation,sharma2013role}.This assessment enabled \trc{} to predict the supervisor's communication style, decision-making preferences, and likely reactions when pushed or under pressure during negotiations, allowing the system to tailor both strategic advice and simulation behaviors to match the specific supervisor's profile.
    
    \para{Theory-guided Advice Generation.}  Third, \trc{} delivered strategic coaching through incremental dialogue building, introducing negotiation concepts gradually across multiple conversation turns rather than providing comprehensive advice all at once. Each turn built on previous exchanges, allowing users to process guidance in manageable pieces while the system adapted advice complexity based on user responses and conversation context.

    \para{Role Based Rehearsal Simulation.} Fourth, \trc{} implemented interactive practice sessions where it embodied the supervisor's personality profile to create realistic but controlled negotiation experiences.
    
    \para{Iterative Feedback Integration.} Fifth, \trc{} enabled users to provide feedback on interaction realism and tone, triggering system recalibration for subsequent responses.
% \end{enumerate}

% This two phase structure, advice generation followed by rehearsal, aimed to operationalize the preparation and practice components required for an effective negotiation coach. 

% \subsubsection{Key Design Features}
\subsubsection{Technical Implementation}
\begin{figure*}[t]
        \centering
        \includegraphics[width=1.99\columnwidth]{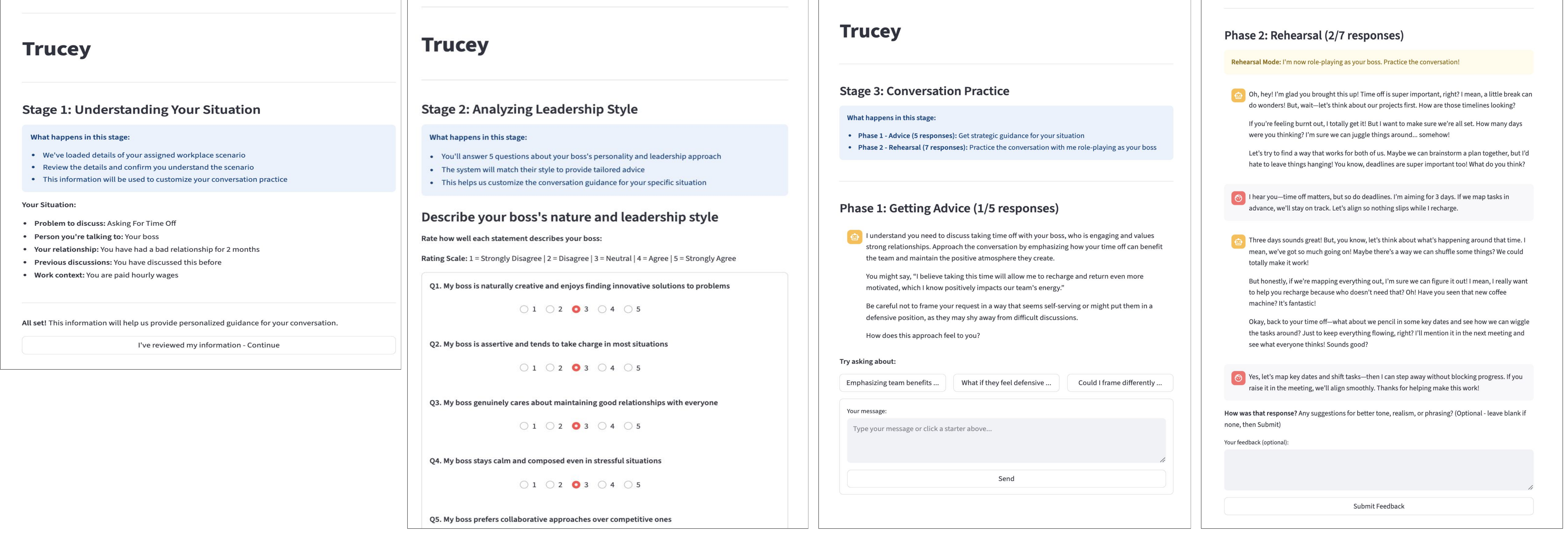}
    \Description[table]{This figure shows a series of screenshots from the Trucey prototype, demonstrating the sequential workflow a user experiences during a negotiation coaching session. The first screenshot, labeled "Stage 1: Understanding Your Situation," give the user their pre-assigned contextual details of a negotiation scenario, such as the problem to discuss, the person they're talking to, and their relationship with that person.
The second screenshot, "Stage 2: Analysing Leadership Style," shows a structured assessment where the user is asked to describe their boss’s nature and leadership style. This assessment, based on Big-Five personality traits, enables the system to tailor its strategic advice and simulation behavior to the specific supervisor's profile.
The third screenshot, "Stage 3: Conversation Practice," is where the core interaction occurs. It is divided into two phases: Phase 1: Getting Advice and Phase 2: Rehearsal. In the "Getting Advice" phase, Trucey provides strategic coaching through incremental dialogue, introducing negotiation concepts gradually to prevent user overwhelm. The "Rehearsal" phase is an interactive practice session where the AI embodies the supervisor's personality profile to create a realistic negotiation experience. 
This phase also includes an "Iterative Feedback Integration" step, which is shown at the bottom of the last screenshot. This feature allows users to provide direct feedback to the chatbot on its realism and tone. This feedback then helps to recalibrate the system for subsequent responses. The goal of this step is to make the coaching system more situationally responsive and to maintain a shared understanding between the user and the AI. The system is technically implemented using GPT-4.1 with a Python Streamlit interface and a SQL database to facilitate these personalized interactions.
}
    \caption{Example screenshots of \trc{} through various stages of user interactions.}
    \label{fig:screenshot}
    % \centering
    % \begin{subfigure}{0.5\columnwidth}
    %     \centering        \includegraphics[width=\columnwidth]{images/Trucey Image 1_v2.pdf}
    %     % \caption{Subfigure 1}
    % \end{subfigure}\hfill
    % \begin{subfigure}{0.5\columnwidth}
    %     \centering
    %     \includegraphics[width=\columnwidth]{images/Trucey Image 2_v2.pdf}
    %     % \caption{Subfigure 2}
    % \end{subfigure}
\end{figure*}

We engineered \trc{} using GPT-4.1 via OpenAI's API, with a Python Streamlit interface and a SQL database for configuration and interaction logging (\autoref{fig:screenshot}). 

% \begin{figure}[t]
%     \centering
%     \begin{subfigure}{0.5\columnwidth}
%         \centering
%         \includegraphics[trim={3cm 0 3cm 0}, clip, width=\textwidth]{images/Untitled Presentation (5).pdf}
%         % \caption{Subfigure 1}
%     \end{subfigure}\hfill
%     \begin{subfigure}{0.5\columnwidth}
%         \centering
%         \includegraphics[trim={3cm 0 3cm 0}, clip, width=\textwidth]{images/Untitled Presentation (4).pdf}
%         % \caption{Subfigure 2}
%     \end{subfigure}
% \end{figure}

To personalize, we implemented a personality-matching mechanism that transformed users' self-reported responses into vector embeddings and used cosine similarity to identify the most relevant supervisor profile. Each profile encoded characteristic strengths and challenges, enabling tailored strategic advice and realistic simulation behaviors---even under simulated pressure scenarios. 

% For a theory-driven strategy, we applied few-shot learning with examples drawn from 
We applied few-shot learning with a theory-driven strategy, drawing examples from~\citeauthor{brett2016negotiation}'s negotiation framework across five difficulty levels. These range from distributive tactics (e.g., ultimatums, direct demands) to integrative strategies (e.g., collaborative framing, interest alignment). 
This approach enables \trc{} to dynamically adjust advice and practice activities according to both user context and the selected supervisor profile.

% The system is implement using GPT-4.1 via Open AI's API, with a python Streamlit interface and SQL Database for system configuration and interaction log storage. 

% \trc{} employs a personality matching system that converts user survey responses into vector embeddings and identifies the most appropriate supervisor behavioral profile using cosine similarity. Each supervisor profile contains specific strengths and challenges, enabling the system to customize both strategic advice and simulation behaviors even under simulated pressure situations. 

% The system's theory-driven prompting approach utilizes few-shot learning with examples derived from Brett \& Thompson's negotiation framework across five difficulty levels. These range from distributive approaches (ultimatums and direct demands) to integrative strategies (collaborative framing and interest alignment), allowing dynamic selection of appropriate theoretical elements based on user context and supervisor personality.

\subsection{Conducting an Experimental Study}
Toward our RQ1 of evaluating the effectiveness of \trc{} as a negotiation coach, we conducted a large-scale online experiment on Prolific, an online academic research platform designed for behavioral research~\cite{palan2018prolific}. 
The study was designed to simulate realistic workplace negotiation contexts and systematically compare across conditions. 
This procedure included participant screening and recruitment, informed consent, compensation, pre- and post- task surveys, and exposure to one of the three study conditions, which we describe below:

% \koustuv{We conducted an experimental study on Prolific. subsubsections should be on participant recruitment, compensation, pre- and post- task surveys, and then the study conditions---}

\subsubsection{Participant Recruitment}
% The participants for the experimental study were recruited through Prolific, an online academic research platform designed for behavioral research. 
We recruited participants through Prolific's default interface. 
To ensure relevance to workplace negotiation contexts, several inclusion criteria were applied. 
Participants were required to be at least 18 years old, proficient in English, and have current or past work experience in a U.S. workplace. 
While U.S. residency was not mandatory, U.S. workplace experience was essential, as cultural norms play a critical role in shaping negotiation strategies~\cite{brett2016negotiation}
Additionally, participants were required to have worked under a supervisor at some point in their career and to have experience initiating negotiations through asynchronous online communication (e.g., text or email).

Interested participants completed an initial screening survey and provided informed consent before proceeding. 
Attention checks were embedded throughout the study procedure, and participants failing these checks or timing out were excluded. 
% assessing these inclusion criteria. Those meeting the requirements provided informed consent before proceeding to the main study. To ensure data quality, participants were excluded if they failed attention checks embedded in the pre- and post-task surveys. 
Of the 368 initial participants, 87 were screened out for not fulfilling study requirements, and 14 failed attention checks or were timed out during the study. 
The remaining 267 participants constituted our final analytical sample. These participants reflected a diverse set of demographic and professional backgrounds (\autoref{tab:demographic_experimental}). Participants who completed the study received \$5 USD in compensation, while those excluded during screening were compensated \$0.15 USD for their time. On average, the study took 26.4 minutes to complete.
These participants completed both pre- and post-task surveys and were randomly assigned to one of the three conditions in approximately a 2:1:1 ratio: \trc{} ($N$=134), \cgpt{}($N$=66), and \hbk{} ($N$=67).

% We conducted an a priori power analysis to determine an adequate sample size for detecting meaningful effects. Following Dattalo's approach for multiple regression analysis, with an anticipated medium effect size=0.15, a desired statistical power level of 0.9, the minimum required sample size was 118 participants. Our achieved sample of $N$=267 exceeds the threshold, providing sufficient statistical power to detect medium to large effects and enabling robust analysis of interaction effects between conditions and negotiation elements.

% a pre- and post- task survey and for the task, were randomly assigned to one of three conditions with roughly a 2:1:1 ratio: \trc{} ($N$=134), \cgpt{}($N$=66), and \hbk{} ($N$=67).

% 
% Our final sample of 267 participants represented diverse demographic and professional backgrounds. Participants who completed the study received \$5 USD Compensation, while those screened out received \$0.15 USD for their time investment. The complete study took on average of 26.4 minutes across all participants.

\begin{table}[t!]
\centering
\sffamily
\footnotesize
   \caption{Demographic distribution of $N$=267 participants in the experimental study.}  
   \label{tab:demographic_experimental}
\setlength{\tabcolsep}{3pt}
\begin{tabular}{lrrrr}
\textbf{} & \textbf{Overall} & \textbf{\trc{}} & \textbf{\cgpt{}} & \textbf{\hbk{}} \\
\toprule
\rowcolmedium \multicolumn{5}{l}{\textit{Sex}}\\
~Male & 141 & 71 & 33 & 37\\
~Female & 121 & 61 & 32 & 28\\
~Other/Prefer Not to Say & 5 & 2 & 1 & 2\\
\rowcolmedium \multicolumn{5}{l}{\textit{Race/Ethnicity}}\\
~White & 181 & 94 & 38 & 49\\
~Black or African American & 50 & 29 & 14 & 7\\
~Hispanic or Latino & 6 & 2 & 3 & 1\\
~Asian & 10 & 2 & 3 & 5\\
~Mixed/Other & 20 & 7 & 8 & 5\\
\rowcolmedium\multicolumn{5}{l}{\textit{Educational Level}}\\
~Associate Degree & 29 & 16 & 7 & 6\\
~Bachelors Degree & 114 & 50 & 29 & 35\\
~Graduate Degree & 80 & 44 & 20 & 16\\
~Other & 44 & 24 & 10 & 10\\
\rowcolmedium \multicolumn{5}{l}{\textit{Employment Type}}\\
~Employed full-time & 197 & 99 & 46 & 52\\
~Employed part-time & 43 & 21 & 10 & 12\\
~Other & 27 & 14 & 10 & 3\\
\rowcolmedium \multicolumn{5}{l}{\textit{Work Domain}}\\
~Administration/Operations & 10 & 5 & 2 & 3\\
~Creative/Design/Marketing & 15 & 5 & 5 & 5\\
~Education/Training & 21 & 14 & 4 & 3\\
~Engineering/Technology/Software & 46 & 24 & 12 & 10\\
~Finance/Accounting/Banking & 24 & 10 & 5 & 9\\
~Government/Public Service & 9 & 2 & 3 & 4\\
~Healthcare/Medical & 26 & 13 & 7 & 6\\
~Legal/Compliance & 4 & 1 & 3 & 0\\
~Management/Leadership & 16 & 8 & 4 & 4\\
~Sales/Business Development & 23 & 17 & 3 & 3\\
~Research/Analytics & 6 & 2 & 1 & 3\\
~Other/Mixed & 67 & 33 & 17 & 17\\
\rowcolmedium \multicolumn{5}{l}{\textit{Compensation Type}}\\
~Salaried (fixed annual amount) & 124 & 62 & 34 & 28\\
~Hourly wages & 81 & 41 & 21 & 19\\
~Other & 62 & 31 & 11 & 20\\
\rowcolmedium \multicolumn{5}{l}{\textit{Work Experience}}\\
~0-1 year & 4 & 1 & 3 & 0\\
~1-3 years & 16 & 9 & 2 & 5\\
~3-5 years & 22 & 10 & 9 & 3\\
~5-7 years & 29 & 14 & 8 & 7\\
~7 years or more & 195 & 100 & 44 & 51\\
~No response & 1 & 0 & 0 & 1\\
\bottomrule
\end{tabular}
\Description[table]{This table provides a detailed demographic breakdown of the 267 participants in the experimental study. It categorizes the participants based on several key characteristics, including Sex, Race/Ethnicity, Educational Level, Employment Type, Work Domain, Compensation Type, and Work Experience. The data is presented for the overall study sample as well as for each of the three intervention conditions: Trucey, ChatGPT, and the traditional Handbook. The table shows the distribution of participants across these demographic and professional categories, ensuring the study's sample reflected a diverse set of backgrounds.
}
\end{table}

% \begin{figure}[H]
%     \centering
%     \includegraphics[width=1.0\textwidth]{visualizations/Participant Flow Chart.png}
%     \caption{Selected Participant Flow Diagram \koustuv{Not sure this figure adds much value---this should be clear in the writing.  }}
%     \label{fig:myfig}
% \end{figure}

\subsubsection{Pre- and Post- Task Surveys}

We employed a set of validated and study-specific measures to assess participants' psychological preparedness for workplace negotiations and their perceptions of the intervention. 
All participants completed identical pre- and post- task surveys to establish baseline and outcome measures, in addition to their task (in assigned condition).%, and finally a post-task survey. 
% followed by post-task surveys after experiencing their assigned condition. 

\para{Pre-Task Measures.} Participants first completed demographic questions and a battery of survey questionnaires on---1) Perceived AI Literacy using the PAILQ-6 questionnaire~\cite{grassini2024psychometric}, 2) Personality Traits using the BFI-10 scale~\cite{raaijmakers1999effectiveness, soto2017next}, a validated 10-item measure of personality traits rated on 5-point Likert scales, 3) Occupational self-efficacy using the Occupational Self-Efficacy-6 Scale (OSS-6)~\cite{rigotti2008short}, and 4) Personal empowerment using the Personal Empowerment Understanding (PEU) questionnaire~\cite{spreitzer1995psychological} measuring four dimensions (meaning, competence, self-determination, and impact), and 5) Negotiation preparedness, measured using two questions on perceived fear for negotiation and the willingness to initiate a negotiation---both using 5-point Likert scales.

% The participants then completed baseline measures of occupational self-efficacy using a modified version of the Occupational Self-Efficacy Scale \cite{rigotti2008short}, psychological empowerment using a modified 9-item version of Spreitzer's Psychological Empowerment Instrument \cite{spreitzer1995psychological} measuring four dimensions (meaning, competence, self-determination, and impact), and fear of negotiation using a custom 5-point Likert scale.

\para{Post-task measures.} After completing the negotiation task, participants completed the same 1) OSS-6~\cite{rigotti2008short}, 2) PEU~\cite{spreitzer1995psychological}, and the 3) negotiation preparedness questionnaires included above. 
In addition, participants rated the intervention's 4) usability using a  4-item Usability Metric for User Experience (UMUX)~\cite{finstad2010usability} and 5) appropriateness using the Intervention Appropriateness Measure (IAM)~\cite{weiner2017psychometric}.
% same occupational self-efficacy, psychological empowerment, and fear of negotiation scales administered pre-task. Additionally, they completed the 4-item Usability Metric for User Experience (UMUX) \cite{finstad2010usability} on 7-point Likert scales and a modified version of the Intervention Appropriateness Measure \cite{weiner2017psychometric} on 5-point Likert scales to assess the appropriateness of the AI intervention.

\subsubsection{Experimental Study Conditions}
% To evaluate the effectiveness of theory-driven AI coaching for workplace negotiations, 
For the task during the experimental study, participants were randomly assigned to one of the three conditions we designed.
All participants worked through one of the three workplace negotiation scenarios---\textit{requesting a raise, a promotion, or time off}. Time on task was held constant across conditions to ensure comparability. 
We designed the three conditions to enable us to disentangle the effectiveness of \trc{} from both generic AI interactions (\cgpt{}) and static theory-based instructions (\hbk{}).
We describe the conditions below:

% and spent similar time in their assigned condition, ensuring comparability across groups. This design allowed us to disentangle the the effectiveness of \trc{} beyond generic AI interaction and from static theory-based instructions.

% ,
% as described below. 
% we designed three conditions
% a three-condition between-subjects experiment that isolated the key components of our approach. The study design aims to discern the individual and interactive effects of AI-mediated interactions and theoretical content.

\para{\trc{} (Interactive AI + Theory-driven resource.) Condition} was the experimental condition, where participants interacted with our prototype, \trc{}. This prototype integrated negotiation theory with personality calibration, contextual advice generation, and an interactive rehearsal simulation. Participants engaged directly with \trc{} to receive theory-driven, tailored coaching and practice strategies for an assigned negotiation scenario, including asking for a raise, asking for a promotion, or asking for time off.

% serves as our experimental condition, implementing a theory-driven AI system with personality calibration, contextual advice generation, and an interactive rehearsal simulation. 
% In this condition, participants needed to interact directly with our prototype \trc{}.
% This experimental condition allows for an assessment of the extent to which an AI coaching tool, when informed by negotiation theory, enhances the efficacy of negotiation preparation.

\para{\cgpt{} (Interactive AI only) Condition} served as the AI control condition. Here, participants interacted with a generic conversational AI (ChatGPT) without theoretical scaffolding or personality calibration. 
To minimize cross-LLM variability, we used GPT-4.5---the same underlying model as in \trc{}.
This condition isolated the potential benefits of AI interaction itself, independent of negotiation theory or personalization mechanisms.

% , allowing users to interact through standard conversation AI without theoretical scaffolding or personality calibration. 
% The condition isolated whether there are any observed benefits that stem from an AI interaction itself or from a theoretically grounded AI and personalization mechanisms.

\para{\hbk{} (Static theory-driven resource) Condition} served as another control condition, consisting of a negotiation resource (without any AI or interactiveness). 
% the theoretical-resource condition.
Participants received the same negotiation content provided in \trc{}---drawn from the same negotiation frameworks~\cite{brett2016negotiation,Bradley_Campbell_2016}---but presented as static, text-based instructions. This condition isolated the contribution of theoretical material from the medium of delivery.

% , delivering identical theoretical material as \trc{}, through a static text based instruction. This experimental condition is designed to isolate whether observed benefits stem from the theoretical content or the interaction medium.

% conditions utilized identical workplace negotiation scenarios and equivalent time commitments to ensure fair comparison. The design enables the following comparisons:
% \begin{enumerate}
%     \item \trc{} vs \cgpt{}: Is there value of theoretical framework integration?
%     \item \trc{} vs \hbk{}: Is there a difference created by an AI-mediated interaction versus the traditional instructions?
% \end{enumerate}

\subsection{Conducting a Semi-structured Interview Study}

To complement the experimental findings and gain deeper insight into participants’ experiences with the interventions, we conducted a semi-structured interview study. Semi-structured interviews are well-suited for capturing nuanced perspectives while still providing consistency across sessions. Our protocol combined a core set of guiding questions with flexibility to probe participants' reflections on their interaction with the conditions, their perceptions of its usefulness, and how it shaped their negotiation preparation.
This approach enabled us to move beyond surface-level outcomes to explore the mechanisms through which participants engaged with the interventions, the challenges they encountered, and the value they attributed to different forms of support.
In this section, we describe our participant recruitment and interview design methodology.
% \koustuv{We conducted an interview study to understand the nuances and deeper insights into how participants navigated through the interventions. Describe your interview methodology}

\subsubsection{Participant Recruitment}

% \begin{table}[t]
% \centering
% \sffamily
% \footnotesize
% \caption{Demographic distribution of $N$=15 participants in the interview study. \koustuv{Could you also add educational level?}}
% \label{tab:qualitative_participants}
% \begin{tabular}{lllll}
% % \toprule
% \textbf{ID} & \textbf{Gender} & \textbf{Race} & \textbf{Work Experience} & \textbf{Work Domain} \\
% \midrule
% P1 & Male & Asian & 1-3 years & Engineering/Technology \\
% P2 & Male & Asian & 0-1 year & Engineering/Technology \\
% P3 & Female & White & 1-3 years & Engineering/Technology \\
% P4 & Male & Asian & 3-5 years & Education \\
% P5 & Male & Asian & 1-3 years & Engineering/Technology \\
% P6 & Female & Asian & 3-5 years & Engineering/Technology \\
% P7 & Female & Asian & 0-1 year & Engineering/Technology \\
% P8 & Male & Asian & 5-7 years & Engineering/Technology \\
% P9 & Male & Asian & 0-1 year & Engineering/Technology \\
% P10 & Female & Asian & 1-3 years & Engineering/Technology \\
% P11 & Male & Asian & 1-3 years & Engineering/Technology \\
% P12 & Female & Asian & 3-5 years & Legal/Management \\
% P13 & Male & Asian & 5-7 years & Engineering/Technology \\
% P14 & Male & Asian & 3-5 years & Engineering/Technology \\
% P15 & Male & Asian & 7+ years & Engineering/Technology \\
% \bottomrule
% \end{tabular}
% \end{table}

\begin{table}[t]
\centering
\sffamily
\footnotesize
\caption{Demographic distribution of $N$=15 participants in the interview study.}
\setlength{\tabcolsep}{3pt}
\label{tab:qualitative_participants}
\begin{tabular}{llllll}
% \toprule
\textbf{ID} & \textbf{Gender} & \textbf{Race} & \textbf{Work Exp.} & \textbf{Work Domain} & \textbf{Education} \\
\midrule
P1 & Male & Asian & 1-3 years & Engineering/Technology & Bachelor's\\
\rowcollight P2 & Male & Asian & 0-1 year & Engineering/Technology & Bachelor \\
P3 & Female & White & 1-3 years & Engineering/Technology & Bachelor\\
\rowcollight P4 & Male & Asian & 3-5 years & Education & Bachelor \\
P5 & Male & Asian & 1-3 years & Engineering/Technology & Graduate \\
\rowcollight P6 & Female & Asian & 3-5 years & Engineering/Technology & Graduate\\
P7 & Female & Asian & 0-1 year & Engineering/Technology & Graduate \\
\rowcollight P8 & Male & Asian & 5-7 years & Engineering/Technology & Graduate\\
P9 & Male & Asian & 0-1 year & Engineering/Technology & Graduate\\
\rowcollight P10 & Female & Asian & 1-3 years & Engineering/Technology & Bachelor\\
P11 & Male & Asian & 1-3 years & Engineering/Technology & Graduate\\
\rowcollight P12 & Female & Asian & 3-5 years & Legal/Management & Graduate \\
P13 & Male & Asian & 5-7 years & Engineering/Technology & Graduate\\
\rowcollight P14 & Male & Asian & 3-5 years & Engineering/Technology & Graduate\\
P15 & Male & Asian & 7+ years & Engineering/Technology & Graduate\\
\bottomrule
\end{tabular}
\Description[table]{This table details the demographics of the 15 participants in the semi-structured interview study. It provides a breakdown of each participant's gender, race, work experience, work domain, and education level.}
\end{table}

We recruited participants for the semi-structured interviews through targeted LinkedIn posts advertising our research on AI coaching for workplace negotiations. Interested individuals completed a Qualtrics screening form, provided consent, and were assessed for eligibility. We applied criteria very similar to the experimental study: participants had to be 18 or older, proficient in English, and have U.S. workplace experience. 
We received 138 responses to our interest form between July and August 2025. 
After screening out for inauthentic or low-quality responses through Qualtrics survey and phone-call-based screening, we recruited 15 participants for our study.
\autoref{tab:qualitative_participants} summarizes the demographics of the participants---while our recruited participants represent a mix of genders and professional backgrounds, with varying levels of workplace experience, the participant pool seemingly lack racial/ethnic diversity, which is a limitation. 
That said, ``Asian'' should not be understood as a single, uniform category---14 of our 15 participants identified as Asian, but represented cultural heterogeneity, including South Asian, East Asian, and Asian American participants, with diversity in immigrant status and countries of origin.
% , underscoring that ``Asian'' should not be understood as a single, uniform category. 
The participants were interviewed for a one-hour remote interview conducted over Teams/Zoom. 
Each interview participant received a \$20 USD Amazon gift card upon completion.

\subsubsection{Interview Design}
% Each interview was a 60-minute session conducted via a Zoom call. It 
Our interviews followed a structured protocol designed to balance standardization with a flexibility for deeper exploration. 
The first four interviews (with P1-P4) focused exclusively on \trc{}, allowing us to probe deeply into its usability, potential issues, and participants' interactions through extended think-aloud exercises and follow-up questioning. Based on these sessions, we revised the protocol to enable comparative evaluation across all three conditions (\trc{}, \cgpt{}, and \hbk{}). 

Accordingly, participants engaged with both AI conditions (\trc{} and \cgpt{}) in a counterbalanced order, spending 20 minutes with each, followed by a 10-minute review of \hbk{}. To mitigate order effects, the sequence of AI conditions was systematically varied across participants. For each AI condition, participants worked through workplace negotiation scenarios using identical prompts to ensure comparability while still allowing for natural variation in responses. 
The interviewer provided minimal guidance during these tasks to capture authentic user interaction.

After each condition, participants answered a series of open-ended questions exploring preferences, perceived effectiveness, and the impact on confidence, efficacy, and negotiation-related anxieties. This structure encouraged detailed, reflective responses while maintaining focus on how the interventions shaped user perceptions.
All interviews were conducted via Zoom, recorded with participant consent, and transcribed. 
% transcribed using AI-assisted transcription with manual verification for accuracy. 
The within-subjects design of interviews enabled direct comparisons across conditions and generated rich insights into the mechanisms underlying our experimental findings.

% longer think-aloud and back-and-forth follow-up questions. 
% After these interviews, we modified the protocol to enable comparative evaluation across all the three conditions (\trc{}, \cgpt{}, and \hbk{}). 
% Participants experienced both AI conditions (\trc{} and \cgpt{}) in a counterbalanced order, spending 20 minutes with each system. This was followed by a 10-minute review of \hbk{}. To control for order effects, the sequence of AI condition presentation across participants was systematically varied. 

% For each AI condition, participants engaged with standardized workplace negotiation scenarios using identical prompts to ensure consistency while allowing natural variation in responses. The interviewer provided minimal guidance during condition experiences to capture authentic user interactions. 

% Following each condition, participants engaged in a set of deep dive questions by exploring their preferences, perceived effectiveness and the impact on their confidence, efficacy and fear primarily. The interview used open-ended questions to encourage detailed responses while maintaining a focus on understanding the impact of this interaction on a user. 

% All sessions were recorded with participant consent and transcribed using AI assistance with manual verification for accuracy. A within subjects design enabled direct comparison and provided rich insights into understanding the mechanisms of our quantitative experimental findings. 

% \subsection{Data Analysis}
\subsection{Privacy, Ethics, and Reflexivity}
Our study was reviewed and approved by the Institutional Review Board (IRB) at our university.
Given the sensitive nature of workplace and interpersonal dynamics that surfaced during the surveys and interviews, we implemented strict privacy and ethical safeguards. 
The experimental study dataset was collected through Prolific and Qualtrics platforms, and was already anonymized at source.
For the interview dataset, we removed all personally identifiable information and paraphrased quotes in the paper to reduce traceability while maintaining context in readership. 

Our interdisciplinary team comprises researchers of diverse gender, racial, and cultural backgrounds, including people of color and immigrants. 
The team brings interdisciplinary expertise in the areas of HCI, CSCW, computational social science, AI ethics, and I/O psychology.
Multiple authors have conducted prior studies on workplace dynamics, and as individuals with lived experience in organizational settings, and our interpretations are also informed by lived experiences of being employed in organizational settings. 
While we have taken the utmost care to capture and faithfully synthesize the participants' viewpoints, we recognize that our interpretations are situated and informed by our disciplinary training, professional backgrounds, and personal experiences. 

%% file: 4findings_rq1.tex
% \section{Results}
\section{RQ1: Psychological Readiness after (AI-driven) Negotiation Coaching}

% \koustuv{Insert the findings from the experimental study. First include the overall findings---the group differences, followed by the regression analyses, and then the linguistic examination of the interactions. }

\subsection{Comparing Psychological Effects of the Interventions}
\autoref{tab:did} summarizes the changes in psychological measures observed in the post-task surveys compared to the pre-task surveys for each of the three interventions---\trc{}, \cgpt{}, and \hbk{}. 
For each comparison between \trc{} and \cgpt{} as well as \trc{} and \hbk{}, we computed effect sizes (Cohen's $d$) and conducted independent-sample $t$-tests for statistical difference. 
Additionally, we performed Kruskal–Wallis $H$-tests to examine overall differences across all three interventions. 
Below, we summarize results for organizational self-efficacy, psychological empowerment, and negotiation preparation (fear and willingness to initiate).

\para{Organizational Self-Efficacy (OSS-6)~\cite{rigotti2008short}.} 
\trc{} led to a slight positive change ($\Delta$=+0.05), while \cgpt{} showed a small decrease ($\Delta$=-0.19) and \hbk{} showed the largest improvement ($\Delta$=+0.58). 
Compared to \cgpt{}, \trc{} performed modestly better, though with a very small effect size ($d$=0.08). 
Relative to \hbk{}, however, \trc{} performed notably worse, with a small negative effect size ($d$=-0.16). 
No statistically significant differences were observed across the three interventions per Kruskal–Wallis test.

\para{Psychological Empowerment (PEU)~\cite{spreitzer1995psychological}.} 
Both \trc{} and \cgpt{} resulted in negligible and similar improvements ($\Delta$=+0.04 each). 
In contrast, \hbk{} yielded a higher increase ($\Delta$=+0.26), with a medium effect size when compared to \trc{} ($d$=-0.40). 
This difference was statistically significant ($t$=-2.63, $p<$0.05), and the Kruskal–Wallis test confirmed significant variation across conditions, indicating that \hbk{} was most effective in enhancing psychological empowerment.

\para{Negotiation Preparedness} 
First, among the three conditions, \trc{} achieved the largest reduction in \textit{negotiation-related fear} ($\Delta$=-0.19). 
In comparison, \cgpt{} increased fear ($\Delta$=+0.14), whereas \hbk{} slightly reduced fear ($\Delta$=-0.10). 
Relative to \cgpt{}, \trc{} achieved a small effect ($d$=-0.27), and compared to \hbk{} the effect was very small ($d$=-0.08). 
Although the changes suggest that \trc{} is promising for alleviating fear.

Next, all three interventions helped increase participants' \textit{willingness to initiate negotiations}. 
\cgpt{} led to the largest improvement ($\Delta$=+0.36), followed by \hbk{} ($\Delta$=+0.31) and \trc{} ($\Delta$=+0.26). 
Effect sizes between interventions were negligible, and the Kruskal–Wallis test showed no significant differences. 
Thus, while willingness to initiate improved across the board, no intervention demonstrated clear superiority. This universal improvement likely reflects that any form of preparation enhances readiness for challenging tasks~\cite{bandura1997self}.

\para{Usability (UMUX)~\cite{finstad2010usability} and Appropriateness (IAM)~\cite{weiner2017psychometric}.} During the exit survey, participants completed the Usability Metric for User Experience (UMUX; 0-100)\cite{finstad2010usability} and the Intervention Appropriateness Measure (IAM; 4-20)\cite{weiner2017psychometric} to evaluate their experience with each coaching condition. The results of these metrics are shown in \autoref{tab:utility}.

For usability, mean UMUX scores were 74.22 for \trc{}, 78.97 for \cgpt{}, and 80.47 for \hbk{}. A Kruskal-Wallis test revealed significant differences across conditions. Pairwise comparisons showed that \trc{} had significantly lower usability ratings than \hbk{} , while the difference between \trc{} and \cgpt{} was not significant. All conditions achieved usability scores above the 68-point threshold for acceptable usability~\cite{finstad2010usability}. This suggests that the \hbk{} was found the easiest to use amongst the users.
For appropriateness, mean IAM scores were 15.63 for \trc{}, 16.11 for \cgpt{}, and 16.45 for \hbk{}, with no significant differences across conditions (. All conditions scored in the \textit{moderately appropriate} to \textit{highly appropriate} range~\cite{weiner2017psychometric}, suggesting participants viewed each coaching approach as a moderately appropriate fit with potential value for negotiation preparation.

% Thus, while willingness to initiate improved across the board, no intervention demonstrated clear superiority.
% \koustuv{This observation could be tied to the fact that practice makes people feel more ready for something~\cite{}.}

% \autoref{tab:did} shows the change in the pscyhological measures in each of the post-task surveys compared to the pre-task suirveys for each of the three intervention condictions---\trc{}, \cgpt{}, and \hbk{}.
% For each pair of \{\trc{}, \cgpt{}\} and \{\trc{}, \hbk{}\}, we obtained effect size (Cohen's $d$) and independent-sample $t$-tests. 
% In addition, we obtained Kruskal-Wallis $H$-test to obtain the statistical significance in differences across the three types of interventions. 

% along with effect size (Cohen's $d$) and indepdent-sample $t$-tests between \trc{} and \cgpt{} and \trc{} and \hbk{}. 

\begin{table*}[t!]
\centering
\sffamily
\footnotesize
   \caption{Comparison of the changes in pre- and post- task surveys.}  
   \label{tab:did}
% \setlength{\tabcolsep}{2pt}
% \begin{tabular}{l|rrr|rrl|rrl|rl}
\begin{tabular}{lrrrcr@{}lcr@{}lr@{}l}
% \textbf{Measure} & \textbf{Trucey} & \textbf{ChatGPT} & \textbf{Handbook} & \multicolumn{3}{c}{\textbf{Trucey v ChatGPT} & \multicolumn{3}{c}{\textbf{Trucey  Handbook}} & \multicolumn{2}{c}{\textbf{Kruskal-Wallis}} \\
\textbf{Measure} & \textbf{\trc{}} & \textbf{\cgpt{}} & \textbf{\hbk{}} & \multicolumn{3}{c}{\textbf{\trc{} vs \cgpt{}}} & \multicolumn{3}{c}{\textbf{\trc{} vs  \hbk{}}} &  \multicolumn{2}{c}{\textbf{Kruskal-Wallis}}\\
& Mean $\Delta$ & Mean $\Delta$  & Mean $\Delta$ & \textbf{Cohen's $d$}  & \multicolumn{2}{c}{\textbf{$t$-test}} & \textbf{Cohen's $d$} & \multicolumn{2}{c}{\textbf{$t$-test}} & \multicolumn{2}{c}{\textbf{$H$-stat.}}\\
\toprule
\rowcollight \multicolumn{12}{l}{\textit{Organizational Self-Efficacy (OSS-6)}}\\
OSS & 0.05 & -0.19 & 0.58 & 0.08 & 0.48 &  & -0.16 & -1.06 &  & 2.37& \\
\rowcollight \multicolumn{12}{l}{\textit{Psychological Empowerment Understanding (PEU)}}\\
PEU: Meaning & 0.05 & 0.10 & 0.36 & -0.06 & -0.41 &  & -0.38 & -2.47 & & 7.06 & *\\
PEU: Competence & -0.02 & -0.02 & 0.25 & 0.00 & 0.01 & & -0.37 & -2.46 & ** & 6.08 & * \\
PEU: Self-determination & 0.03 & 0.00 & 0.11 & 0.05 & 0.3 &  & -0.13 & -0.81 &  & 2.34 & \\
PEU: Impact & 0.11 & 0.05 & 0.31 & 0.07 & 0.46 &  & -0.28 & -1.81&* & 4.37 & \\
PEU: Aggregated & 0.04 & 0.04 & 0.26 & 0.01 & 0.09 &  & -0.40 & -2.63 & **  & 8.36 & **\\
\hdashline
\rowcollight \multicolumn{12}{l}{\textit{Negotiation Preparedness}}\\ 
Negotiation Fear & -0.19 & 0.14 & -0.1 & -0.27 & -1.8 & * & -0.08 & -0.49 &  & 3.81 & *\\
Negotiation Initiativeness & 0.26 & 0.36 & 0.31 & -0.11 & -0.73 &  & -0.06 & -0.4 &  & 0.75 & \\
\bottomrule
\end{tabular}
\Description[table]{}
\end{table*}

\begin{table*}[t!]
\centering
\sffamily
\footnotesize
   \caption{Comparison of the usability and acceptability of the various conditions of negotiation coaching.}  
   \label{tab:utility}
\begin{tabular}{lrrrcr@{}lcr@{}lr@{}l}
\textbf{Measure} & \textbf{\trc{}} & \textbf{\cgpt{}} & \textbf{\hbk{}} & \multicolumn{3}{c}{\textbf{\trc{} vs \cgpt{}}} & \multicolumn{3}{c}{\textbf{\trc{} vs  \hbk{}}} &  \multicolumn{2}{c}{\textbf{Kruskal-Wallis}}\\
& \textbf{Mean} & \textbf{Mean}  & \textbf{Mean} & \textbf{Cohen's $d$}  & \multicolumn{2}{c}{\textbf{$t$-test}} & \textbf{Cohen's $d$} & \multicolumn{2}{c}{\textbf{$t$-test}} & \multicolumn{2}{c}{\textbf{$H$-stat.}}\\
\toprule
UMUX & 74.22 & 78.97 & 80.47 & -0.24 & -1.52 &  & -0.32 & -2.04 & * & 3.19 & **\\
IAM & 15.63 & 16.11 & 16.45 & -0.14 & -0.91 &  & -0.24 & -1.54 &  & 2.53 & \\
\bottomrule
\end{tabular}
\Description[table]{}
\end{table*}

\subsection{Linguistic Analysis}

An interesting observation from the above concerns how the effects of interacting with \trc{} and \cgpt{} did not show statistically significant differences for a majority of outcomes. To unpack this further, we quantitatively examined the differences in the language of interactions by \trc{} and \cgpt{}.

\subsubsection{Linguistic Similarities with Theory-Driven Negotiation Elements}
To begin with, we operationalized and measured the presence of various \citeauthor{brett2016negotiation}'s elements in the language of \trc{} and \cgpt{}~\cite{brett2016negotiation}. 
Our methodological approach is inspired by prior work on examining linguistic markers in workplace-related communication~\cite{saha2019libra,dasswain2020culture}.
% \koustuv{Describe how this analysis was done.}
For this purpose, we employed vector embeddings using Bidirectional Encoder Representations from Transformers (or BERT) language model, using pre-trained BERT-based sentence transformers~\cite{reimers2019sentence}. 
We constructed embeddings for the seven core negotiation concepts in the \citeauthor{brett2016negotiation}'s framework: Strategy Development (SD), Information Asymmetry (IA), Interest Exploration (IE), Outcome Analysis (OA), Long-term Relationship (LTR), and Power Dynamics (PD).
These concepts featured the same theoretical examples used in \trc{}'s few-shot prompting. 
Similarly, we obtained the BERT embeddings of the linguistic interactions in \trc{} and \cgpt{}. 
We then computed cosine similarity scores between AI (\trc{} and \cgpt{}) language in responses and each negotiation element to quantify the degree to which the element was reflected in the responses.
% to measure the degree of \citeauthor{brett2016negotiation}'s aspects in \trc{} and \cgpt{} interactions.

\autoref{tab:brett_elements} summarizes the comparative results across the two AI tools, including relative differences, effect sizes (Cohen's $d$), and $t$-tests. 
Our analysis revealed several key differences. In Strategy Development, \trc{} responses aligned more strongly with most components, except for phased SD. For Information Asymmetry (IA), \trc{} produced slightly fewer references to prepared IA (–2\%) but more strategic IA (+5\%). In Interest Exploration (IE), \trc{} showed reduced creative and limited IE but emphasized mutual and evaluative IE, pointing toward a more collaborative style. Within Outcome Analysis (OA), \trc{} generated fewer flexible (–5\%) and procedural (–1\%) OA elements, though both AI tools were nearly identical on holistic OA. For Long-term Relationships, the two systems performed similarly, indicating parity in addressing relational considerations. Finally, in Power Dynamics, \trc{} highlighted collaborative PD (+5\%) while reducing authority-based PD (–4\%).

Together, these findings suggest that \trc{} is more directive and collaborative, embedding strategy and mutuality at the core of its language, while \cgpt{} remains broader and more balanced across negotiation elements. Notably, the analysis also demonstrates that \cgpt{}---even without specialized prompting---automatically incorporates several of Brett’s components by default, reflecting how large language models may implicitly encode negotiation theories. This distinction highlights the potential of targeted prompting to guide AI systems toward specific negotiation orientations, such as collaboration and strategic planning, beyond the general-purpose balance evident in default outputs.

\begin{table}[t]
\centering
\sffamily
\footnotesize
\caption{Comparison of \citeauthor{brett2016negotiation}'s elements of successful negotiation between \trc{} and \cgpt{} with effect sizes reported as Cohen's d and statistical significance via t-tests.}
\label{tab:brett_elements}
\begin{tabular}{lrrrrr@{}l}
\textbf{Element} & \textbf{\trc{}} & \textbf{\cgpt{}} & \textbf{Mean $\Delta$\%} & \textbf{Cohen's d} & \textbf{t-test}&\\
\toprule
\rowcollight \multicolumn{7}{l}{\textit{Strategy Development}}\\
Basic & 0.45 & 0.48 & -2.38 & -0.26 & -3.60& *** \\
Phased & 0.69 & 0.68 & 1.17 & 0.18 & 2.42 & **\\
Structured & 0.76 & 0.78 & -2.10 & -0.32 & -4.30 & *** \\
Innovative & 0.63 & 0.66 & -2.82 & -0.46 & -6.24& *** \\
\rowcollight \multicolumn{7}{l}{\textit{Information Asymmetry}}\\
Prepared & 0.68 & 0.70 & -1.90 & -0.27 & -3.67& *** \\
Strategic & 0.75 & 0.70 & 5.43 & 0.70 & 9.61& *** \\
\rowcollight\multicolumn{7}{l}{\textit{Interest Exploration}}\\
Limited & 0.35 & 0.40 & -5.03 & -0.53 & -7.27& *** \\
Evaluative & 0.66 & 0.65 & 1.29 & 0.13 & 1.76 & \\
Mutual & 0.69 & 0.68 & 1.17 & 0.18 & 2.42& *** \\
Creative & 0.67 & 0.71 & -3.86 & -0.64 & -8.78& *** \\
\rowcollight\multicolumn{7}{l}{\textit{Outcome Analysis}}\\
Procedural & 0.60 & 0.61 & -1.15 & -0.19 & -2.65 & **\\
Flexible & 0.63 & 0.68 & -5.04 & -0.79 & -10.72& *** \\
Holistic & 0.64 & 0.64 & 0.73 & 0.12 & 1.63 & \\
\rowcollight\multicolumn{7}{l}{\textit{Long-Term Relationship}}\\
Conventional & 0.60 & 0.61 & -1.15 & -0.19 & -2.65 & **\\
Strategic & 0.65 & 0.64 & 0.44 & 0.05 & 0.62 & \\
\rowcollight\multicolumn{7}{l}{\textit{Power Dynamics}}\\
Collaborative & 0.75 & 0.70 & 5.43 & 0.70 & 9.61 & ***\\
Authority Based & 0.43 & 0.47 & -3.96 & -0.41 & -5.59 & ***\\
\bottomrule
\end{tabular}
\Description[table]{This table compares the linguistic alignment of Trucey and ChatGPT with specific negotiation concepts from the Brett and Thompson framework. The analysis shows that Trucey's responses are more closely aligned with a collaborative and strategic approach, while ChatGPT's responses are more general and less focused on power dynamics. Specifically, Trucey's language emphasizes strategic information exchange and mutual interest exploration , whereas ChatGPT's language is broader and less targeted. This highlights how theory-driven prompting can guide an AI toward a specific negotiation style.
}
\end{table}

\begin{table*}[t]
\sffamily
\centering
\footnotesize
\caption{Linear regression models revealing the relationship between individual differences and linguistic similarity of interactions with \citeauthor{brett2016negotiation}'s framework's aspects with our outcome measures (* \n{p}\textless{}0.05, ** \n{p}\textless0.01, *** \n{p}\textless{}0.001). This model only includes the data from the two AI (\trc{} and \cgpt{}) interactions.}
\begin{tabular}{l r@{}l r@{}l r@{}l r@{}l}
    \textbf{Dep. Variables} $\rightarrow{}$ & \multicolumn{2}{c}{\textbf{OSS}} & \multicolumn{2}{c}{\textbf{PEU}} & \multicolumn{2}{c}{\textbf{Fear}} & \multicolumn{2}{c}{\textbf{Initiativeness}}\\ 
    \toprule
    % \textbf{Indep. Variables} $\downarrow{}$ & \textbf{Coeff.} & & \textbf{Coeff.} &  &\textbf{Coeff.} &  &\textbf{Coeff.} &  & &\\  

\textbf{Indep. Variables} $\downarrow{}$ & \textbf{Coeff.} &  & \textbf{Coeff.} &  & \textbf{Coeff.} &  & \textbf{Coeff.} & \\
     \cmidrule(lr){1-1}\cmidrule(lr){2-3}\cmidrule(lr){4-5}\cmidrule(lr){6-7}\cmidrule(lr){8-9} 
% Intercept &  &  &  &  &  &  &  & \\
Gender &  &  &  &  &  &  &  & \\
Race &  &  &  &  &  &  &  & \\
Education &  &  &  &  &  &  &  & \\
Employment Type &  &  &  &  &  &  &  & \\
Work Sector &  &  &  &  &  &  &  & \\
Compensation Type &  &  &  &  &  &  &  & \\
Relationship with Supervisor &  &  &  &  &  &  &  & \\
Condition: Trucey & 6.89 & * & 2.96 &  & -20.43 & ** & 12.65 & *\\
\hdashline
% Strategy Development (SD) &  &  &  &  &  &  &  & \\
% SD: Basic &  &  &  &  &  &  &  & \\
Strategy Development: Basic X Trucey & 520.50 & ** & 885.00 & ** & 115.70 & * &  & \\
Strategy Development: Phased &  &  &  &  & 16.03 & *** & -5.15 & *\\
Strategy Development: Phased X Trucey & 31.61 & * &  &  & -14.89 & ** &  & \\
Strategy Development: Structured &  &  &  &  & -52.34 & * &  & \\
Strategy Development: Structured X Trucey &  &  &  &  &  &  &  & \\
Strategy Development: Innovative & 235.50 & ** & 41.05 & ** & -60.39 & ** & 35.81 & *\\
Strategy Development: Innovative X Trucey & -468.69 & *** & -64.69 & ** &  &  &  & \\
% Information Asymmetry (IA) &  &  &  &  &  &  &  & \\
% Information Asymmetry: Prepared &  &  &  &  &  &  &  & \\
\hdashline
Information Asymmetry: Prepared X Trucey & -109.83 & * &  &  & -27.51 & * &  & \\
Information Asymmetry: Strategic & -25.43 & * & -5.32 & * &  &  &  & \\
Information Asymmetry: Strategic X Trucey & 58.47 & *** & 12.31 & *** &  &  &  & \\
% Interest Exploration (IE) &  &  &  &  &  &  &  & \\
\hdashline
Interest Exploration: Limited & -159.90 & * &  &  &  &  &  & \\
Interest Exploration: Limited X Trucey & 511.80 & ** & 87.02 & ** &  &  &  & \\
Interest Exploration: Evaluative &  &  & -23.00 & * &  &  &  & \\
Interest Exploration: Evaluative X Trucey &  &  & 32.82 & ** & -50.89 & * &  & \\
Interest Exploration: Mutual &  &  &  &  &  &  & -5.16 & *\\
Interest Exploration: Mutual X Trucey &  &  &  &  & -14.80 & * &  & \\
Interest Exploration: Creative &  &  & 13.77 & * & -25.18 & * & 38.51 & *\\
Interest Exploration: Creative X Trucey & -351.67 & *** & -51.33 & *** &  &  &  & \\
% Outcome Analysis &  &  &  &  &  &  &  & \\
\hdashline
Outcome Analysis: Procedural & 33.04 & ** & 7.58 & * & -11.80 & * &  & \\
Outcome Analysis: Procedural X Trucey & -64.65 & ** & -11.96 & ** & 15.35 & * &  & \\
Outcome Analysis: Flexible &  &  &  &  &  &  &  & \\
Outcome Analysis: Flexible X Trucey & 603.65 & *** & 97.86 & ** &  &  &  & \\
Outcome Analysis: Holistic &  &  &  &  &  &  & -34.37 & *\\
Outcome Analysis: Holistic X Trucey & 208.56 & ** &  &  &  &  &  & \\
% Long-Term Relationship &  &  &  &  &  &  &  & \\
\hdashline
Long-Term Relationship: Conventional & 33.30 & * & 7.62 & * & -11.74 & ** &  & \\
Long-Term Relationship: Conventional X Trucey & -64.18 & ** & -11.89 & ** & 15.45 & ** &  & \\
Long-Term Relationship: Strategic & 101.66 & ** & 32.44 & ** &  &  &  & \\
Long-Term Relationship: Strategic X Trucey & -217.46 & *** & -49.27 & *** & 58.53 & ** &  & \\
% Power Dynamics (PD) &  &  &  &  &  &  &  & \\
\hdashline
Power Dynamics: Collaborative & -24.50 & * & -5.17 & * &  &  &  & \\
Power Dynamics: Collaborative X Trucey & 58.24 & *** & 12.28 & *** &  &  &  & \\
% PD: Authority &  &  &  &  &  &  &  & \\
Power Dynamics: Authority X Trucey & -963.40 & ** & -163.80 & * & -214.10 & * &  & \\
\hdashline
% Personality Traits (BFI) &  &  &  &  &  &  &  & \\
BFI: Extraversion &  &  &  &  & 0.24 & * &  & \\
BFI: Agreeableness &  &  &  &  &  &  &  & \\
BFI: Conscientiousness &  &  & 0.01 & * & -0.26 & * & 0.21 & *\\
BFI: Neuroticism &  &  &  &  & 0.09 & * &  & \\
BFI: Openness &  &  &  &  &  &  &  & \\
AI Literacy &  &  &  &  &  &  &  & \\
OSS: Pre-task & 0.48 & *** &  &  &  &  &  & \\
PEU: Pre-task & 1.93 & *** & 0.99 & *** &  &  &  & \\
Fear: Pre-task &  &  &  &  & 0.24 & *** &  & \\
Initiativeness: Pre-task &  &  &  &  &  &  & 0.63 & ***\\
Task time &  &  &  &  &  &  &  & \\
% \rowcollight \multicolumn{11}{l}{\textit{Traits}}\\
% ~PANAS: P. Affect & 1.80 &  & 1.25 &  & 1.58 &  & -0.55 &  & \textbf{-6.48} & \textbf{*} & -2.40& \\
% ~PANAS: N. Affect &  -1.07 &  & -3.13 &  & -0.46 &  & -0.08 &  & 1.72 &  & -3.01& \\
\rowcollight & \multicolumn{2}{l}{\n{R^2}=\textbf{0.87***}} &\multicolumn{2}{l}{\n{R^2}=0.86***} & \multicolumn{2}{l}{\n{R^2}=\textbf{0.65**}} & \multicolumn{2}{l}{\n{R^2}=\textbf{0.74***}} \\
    \bottomrule
    \end{tabular}
\label{table:regression}
\Description[table]{This table presents the results of linear regression models that analyze the effects of AI interventions and linguistic patterns on four key outcomes: Occupational Self-Efficacy (OSS), Psychological Empowerment Understanding (PEU), Negotiation Fear, and Initiativeness. The table shows the coefficients for various independent variables, including demographic traits, personality types, and the interaction effects between the Trucey condition and specific negotiation elements. This analysis, which only includes data from the Trucey and ChatGPT interactions, helps to reveal which factors are most predictive of psychological preparedness and how they are moderated by the theory-driven nature of Trucey.}
\end{table*}

\subsubsection{Regression Modeling}
\label{sec: Regression Modeling}
Next, we built linear regression models to examine the effects of AI interventions, demographic and personality covariates, and negotiation element similarities on four outcome measures---occupational self-efficacy (OSS), psychological empowerment understanding (PEU), fear, and willingness to initiate negotiation (initiativeness). 
The models incorporated a rich set of independent variables collected during the pre-task survey, including demographic attributes (e.g., gender, race, education), workplace-related variables (e.g., employment type, work sector compensation type, relationship with supervisor), personality traits (BFI: extraversion, agreeableness, conscientiousness, neuroticism, openness), AI literacy, and pre-task baselines of the outcome measures themselves (e.g., pre-task OSS, PEU, fear, initiativeness).
In addition, we tested a range of models with interaction terms. While several combinations were explored, the most robust and significant patterns emerged when including interaction effects between the intervention condition (\trc{} vs. \cgpt{}) and the linguistic similarities to Brett's negotiation elements (reported in the previous section). These interaction terms allowed us to assess not only whether a given negotiation element was associated with the outcomes, but also whether its effect was moderated by the AI.
%condition.

We built separate regression models for each dependent variable, thereby allowing us to isolate the predictors most relevant to each outcome domain. This approach also facilitated a clearer interpretation of how \trc{} and \cgpt{} shaped different facets of negotiation experiences. \autoref{table:regression} summarizes the $\beta$ coefficients and their statistical significance across all models. Below, we unpack the findings for each outcome variable in turn.

Overall, demographic and workplace attributes were not significant across any of the models, and personality traits only appeared sporadically. Instead, the most consistent predictors were the AI intervention condition and interaction terms with negotiation elements.
We unpack these observations below:
% and outcome measures. We used the demographic attributes, personality traits, AI literacy, and other pre-task measures (e.g., OSS, PEU, and negotiation preparedness) as independent variables collected pre-task. 
% We built several models including all combinations of interaction effects. However, the  interaction effect terms were most significant when combined with similarities with negotiation elements (reported above).
% We built separate regression models for each outcome measure as dependent variable---occupational self-efficacy, psychological empowerment understanding, fear, and willingness to initiate negotiation. 
% ~\autoref{table:regression} summarizes the $\beta$ coefficients and significance of the 
% independent variables for each of the outcome measures.
% We describe our observations below.

% None of the demographic and workplace related variables were significant. We find that being subjected to \trc{} condition has a significant positive relationship ($\beta$=6.89) with OSS and initiativeness ($\beta$=12.65), but a negative relationship with fear ($\beta$=20.43). 

\para{Occupational Self-Efficacy (OSS).} Being in the \trc{} condition significantly increased OSS ($\beta$=6.89). Within negotiation elements, basic SD \n{X} \trc{} ($\beta$=520.50) and phased SD \n{X} \trc{} ($\beta$=31.61) were positively associated with OSS, suggesting that structured and stepwise strategy development enhanced self-efficacy when supported by \trc{}.  \trc{}'s dynamic contextual layering works optimally with structured, straightforward approaches as they can be broken down effectively into digestible incremental pieces. However, innovative SD alone was positively associated ($\beta$=235.50), but its interaction with \trc{} showed a significant negative effect ($\beta$=–468.69), indicating that \trc{}'s incremental building approach contradicts the simultaneous understanding required for integrated thinking in creative solutions. Therefore, in the presence of this approach, the system's inherent contradiction undermines a user's self-perceived efficacy in ideating for a negotiation. Negative associations also emerged with limited IE ($\beta$=–159.90) unless combined with \trc{} ($\beta$=511.80), where the direction reversed. \trc{} helps provide users with more negotiation options to consider when they exist in a limited bargaining space due to the nature of their suggestions. Therefore, users feel less helpless in constrained situations. Pre-task OSS strongly predicted post-task OSS ($\beta$=0.48). %The model explained substantial variance ($R^2$=0.87).

% \para{Psychological Empowerment Understanding (PEU).}

\para{Psychological Empowerment Understanding (PEU).} Here again, positive coefficients indicate factors associated with higher PEU, while negative coefficients indicate factors associated with lower PEU. Although \trc{} itself was not a significant predictor, several interaction effects emerged. Basic SD \n{X} \trc{} ($\beta$=885.00) showed a positive effect, demonstrated substantial positive associations with empowerment, suggesting that Trucey's incremental guidance approach enhances users' sense of understanding and control when applied to straightforward negotiation scenarios. The system's structured methodology appears to provide clear pathways for empowerment in contexts where step-by-step building can be effectively integrated. However, while Innovative SD's ($\beta$=41.05), it's interaction with \trc{} was negative ($\beta$=–64.69). This indicated that the system's structured approach may constrain the cognitive flexibility that users associate with empowerment in innovative contexts. Similarly, Creative IE also showed a modest positive association ($\beta$=13.77), which flipped to a large negative interaction under \trc{} ($\beta$=–51.33), suggesting that when users attempt creative approaches, the incremental methodology conflicts with their sense of empowered exploration.
Conversely, Evaluative IE was negatively associated ($\beta$=–23.00), though its interaction with \trc{} reversed to positive ($\beta$=32.82). \trc{} helps provide users with solutions and reminders in approaching situations with ambiguity and therefore pro-actively empower them to react to a situation. Finally, procedural OA and conventional LTR were both positively associated with the outcome variables; however, their interactions with \trc{} were negative. This negative moderation can be attributed to the embodiment of \trc{} elements during simulation, which subsequently diminishes users' sense of empowerment through deferral mechanisms. Similarly, strategic LTR demonstrated a positive main effect that was negatively moderated by \trc{}. This interaction pattern reflects the influence of organizational knowledge disclosure and normative compliance requests, which collectively reduce user empowerment. Understandably, pre-task PEU predicted post-task PEU ($\beta$=0.99). 

\para{Fear of Negotiation.} Here, negative coefficients indicate factors associated with lowering negotiation fear, while positive coefficients indicate factors associated with higher fear. Fear decreased significantly under the \trc{} condition ($\beta$=–20.43).  This reflects how structured AI guidanced and contextual support may reduce  uncertainty-driven negotiation anxiety by making users feel more prepared and supported.Several negotiation elements also shaped fear: phased SD ($\beta$=16.03) was associated with higher fear increasing anxiety by creating anticipation across multiple sequential stages, transforming a single anxiety-provoking event into multiple interconnected worries, whereas structured SD  ($\beta$=–52.34) and innovative SD ($\beta$=–60.39) lowered fear by providing providing clear frameworks. These approaches reduce uncertainty through predictable pathways, reducing fear of rejection or inadequate performance.

For IE, evaluative IE \n{X} \trc{} ($\beta$=–50.89), mutual IE \n{X} \trc{} ($\beta$=–14.80), and creative IE ($\beta$=–25.18, p<.05) reduced fear as \trc{}'s incremental guidance provides scaffolding for complex interpersonal dynamics, making inherently uncertain social processes feel more manageable through structured breakdown and contextual calibration. Conversely, Procedural OA \n{X} \trc{} ($\beta$=15.35) and conventional LTR \n{X} \trc{} ($\beta$=15.45) were associated with greater fear, suggesting that procedural and conventional frames on outcome analysis may feel constraining when paired with \trc{}. Personality traits such as extraversion ($\beta$=0.24), conscientiousness ($\beta$=-0.26), and neuroticism ($\beta$=0.09) showed modest associations. Unsurprisingly, pre-task fear was a strong predictor ($\beta$=0.24).
% Within the regression model predicting OSS, we find that being subjected to \trc{} bears a significant positive relationship ($\beta$=6.89). Next, within the negotiation elements, basic SD when offered by \trc{} has significant positive association ($\beta$=520.50), phased SD interaction with \trc{} has significant positive association ($\beta$=31.61). Interastingly, innovative SD has a positive association ($\beta$=235.50), but when interacting with \trc{} has a significant negative relationship ($\beta$=-468.69). 
% Within information asymmetry, 

\para{Willingness to Initiate Negotiation.} In the case of willingness to initiate negotiations, positive coefficients indicate factors associated with higher initiativeness, and negative coefficients are associated with lower initiativenes. 
We find that initiativeness was significantly higher in the \trc{} condition ($\beta$=12.65). This reflects how structured guidance and support increase users' confidence to initiate negotiations by making them feel better prepared and supported. Innovative SD ($\beta$=35.81) and creative IE ($\beta$=38.51) increased initiativeness. This could be accounted to allowing users to feel like they could produce unique value proposition, increasing their motivation to initiate negotiations. While mutual IE ($\beta$=–5.16),  reduced initiativeness, likely because mutual approaches require consensus from the other party, making users hesitant to initiate, as they perceive a joint agreement is required. Holistic OA ($\beta$=–34.37) reduced initiativeness, suggesting that a comprehensive outcome analysis could create analysis paralysis due to the complexity of factors involved in turn with potential outcomes, making the process of initiating a negotiation overwhelming to start with. Pre-task initiativeness also significantly predicted post-task scores ($\beta$=0.63). %The model explained considerable variance ($R^2$=0.74).

% Next, within the negotiation element similarities,  we find that structured SD 

\subsubsection{Lexico-Semantic Evaluations}
Next, we conducted lexico-semantic comparisons in the interactions with \trc{} and \cgpt{}. We drew on prior work~\cite{saha2025ai,dasswain2025ai} to operationalize and measure several lexico-semantic measures. 
\autoref{tab:lexico_semantics} summarizes the differences in AI's language as well as the users' language in interactions with the AIs. We describe the operationalizations and observations below:

\para{Verbosity.} The quantity of language, or verbosity, has been linked to perceived quality and emotional depth in support exchanges~\cite{glass1992quality,saha2020causal}. We operationalized verbosity as the total word count per interaction turn. 
We find that \trc{}'s responses were 168\% less verbose compared to \cgpt{}'s responses with a very large effect size (Cohen's $d$=-4.14) and statistical significance ($t$=27.55, $p$<0.05). 
Similar pattern was also followed in the users' responses to the AI, with queries to \trc{} were 20\% less verbose in comparison to those to \cgpt{} (Cohen's $d$=-0.37).

\para{Readability.}
Readability reflects the ease with which text can be comprehended~\cite{wang2013assessing,mcinnes2011readability}. We used the Fleisch-Kincaid Readability Ease~\cite{flesch2007flesch}, which calculates readability based on the formula: 206.835 - 1.015 ($\frac{\text{total words}}{\text{total sentence}}$) - 84.6 ($\frac{\text{total syllables}}{\text{total words}}$)

% \noindent{\small
% Flesch-Kincaid Readability Ease = 206.835 - 1.015 ($\frac{\text{total words}}{\text{total sentence}}$) - 84.6 ($\frac{\text{total syllables}}{\text{total words}}$)
% }

Higher readability ease indicates that the writing is easier to read.
\trc{} responses showed significantly higher readability scores (\trc{} mean = 80.00) than \cgpt{} responses (mean = 68.46), with a very large effect size (Cohen's $d$=2.17). 
This means that \trc{}'s responses had lower linguistic sophistication than \cgpt{}'s responses, and were easier to comprehend with lower levels of English education. 
% higher linguistic sophistication in \cgpt{}'s responses, it may pose barriers for individuals with lower English language comprehension levels. 
That said, the language of participants' queries to both \trc{} and \cgpt{} showed no significant differences. 

\para{Repeatability.}
Repeatability, operationalized as the normalized frequency of non-unique words per sentence, serves as a measure of redundancy in written expression~\cite{ernala2017linguistic,saha2018social}. 
Although repetition may reinforce key ideas, excessive use of the same words can reflect reduced linguistic precision. 
\trc{} responses showed significantly lower (-8.55\%) repeatability compared to \cgpt{} responses with large effect size (Cohen's $d$=-1.17). 
This may be aligned with the notion that \trc{} was more specific and specialized to negotiation rehearsal, whereas \cgpt{}, being more generic, showed tendency toward reiterative phrasing. 
Interestingly, the language of users' queries showed an opposite pattern---users showed 11\% more repetitive language in their queries to \trc{} than to \cgpt{}.

\para{Complexity.}
Linguistic complexity captures the degree of syntactic and lexical sophistication and has been linked to cognitive effort in communication~\cite{kolden2011congruence}. We quantified complexity using the average length of words per sentence~\cite{ernala2017linguistic}. 
We find that \trc{}'s responses were 2.84\% less complex compared to \cgpt{}'s responses with a large effect size (Cohen's $d$=-0.67). 
The users' queries to \trc{} and \cgpt{} did not show any statistically significant difference.

\para{Categorical-Dynamic Index (CDI).}
Language style can be conceptualized along a spectrum from categorical (structured and analytic) to dynamic (personal and narrative-driven)~\citep{pennebaker2014small}. The Categorical-Dynamic Index (CDI) captures this continuum using the formula:

\noindent{\small
CDI = (30 + \text{articles} + \text{prepositions} - \text{personal pronouns} - \text{impersonal pronouns} - \text{auxiliary verbs} - \text{conjunctions} - \text{adverbs} - \text{negations}).
}

Higher CDI indicates a categorical style of writing, and a lower CDI indicates a dynamic or narrative style of writing. 
We computed CDI scores using LIWC-derived part-of-speech frequencies~\citep{tausczik2010psychological}. 
Interestingly \trc{} showed a significantly lower CDI (by -30\%) compared to \cgpt{} with a large effect size ($d$=-0.76). 
This indicates that \trc{}'s language was significantly more dynamic and conversational in style than \cgpt{}. 
However, users' queries to both \trc{} and \cgpt{} did not show any statistical significance. 

\para{Politeness.} Politeness contributes to rapport and trust in therapeutic and peer-support conversations~\cite{brown1987politeness}. We used a pre-trained classifier~\cite{srinivasan2022tydip} that assigns a politeness score between 0 and 1. \trc{} responses were significantly more polite (by 1.55\%) than \cgpt{} responses with medium effect size (Cohen's $d$=0.56).
However, the language of users' queries to both the AIs did not show any significant differences.

\para{Formality.} Formality reflects sociolinguistic variation and has been linked to contextual appropriateness and audience expectations~\cite{heylighen1999formality,larsson2020syntactic}. We employed a RoBERTa-based classifier~\cite{babakov2023don} trained on the GYAFC (Grammarly's Yahoo Answers Formality Corpus) dataset~\cite{rao2018dear,pavlick2016empirical}, which assigns a probability score (0 to 1) reflecting linguistic formality. 
\trc{} responses showed a significantly lower (by -21\%) formality than \cgpt{} responses with very large effect size (Cohen's $d$=-1.73). 
This suggests that \trc{}'s tone was more casual---plausibly due to our fine-tuning by incorporating negotiation framework elements~\cite{brett2016negotiation}. 
% lower formality than \oc{} responses (GPT $d$ = -0.21), suggesting a more casual and accessible tone.

\para{Empathy.}
Empathy is central to supportive communication and is defined as the expression of understanding, validation, and emotional resonance~\cite{herlin2016dimensions,sharma2020computational}. We used a RoBERTa-based empathy model fine-tuned on responses to emotionally evocative content~\cite{buechel2018modeling,tafreshi2021wassa}. 
Here, \trc{} responses showed significantly lower (by -11.36\%) lower empathy than \cgpt{} responses with large effect size (Cohen's $d$=-1.35). 
Recent research has shown how LLms are increasingly capable of simulating empathetic language~\cite{inzlicht2023praise,kidder2024empathy,welivita2024chatgpt}---however, while incorporating a theory-driven negotiation approach, we did not explicitly fine-tune (or prompt) \trc{} to be empathetic to users' queries.
% That said, the users' queries to both the AIs did not show any statistically significant differences. 
This lower empathy can also be tied to the design choice in \trc{}, where the simulated supervisor was intentionally configured to act mean and resistant in order to create a challenging negotiation context. That said, the users' queries to both AIs did not show any statistically significant differences.
% \koustuv{This can be tied to the fact that \trc{} was designed in a way so that the supervisor was mean in the interaction. }

\para{Persuasiveness.}
Persuasiveness is central to encouraging behavior change and fostering motivation in support-oriented discourse~\citep{tan2016winning}. 
We applied a pre-trained persuasiveness classifier~\citep{wang2019persuasion} that scores text on a continuous scale from 0 to 1. 
We found that \trc{} responses showed higher persuasiveness (by 2\%) than \cgpt{} responses with medium effect size (Cohen's $d$=0.66). 
This may reflect that \trc{} was likely more assertive than \cgpt{}, which may have exhibited more neutral, generic, and cautious language. 
In contrast, the persuasiveness exhibited in users' queries to both AIs did not show any significant differences.

Theory-driven AI coaching produces context-dependent psychological benefits that differ from \cgpt{} and \hbk{}. \trc{} most effectively reduced negotiation fear through accessible, theory-integrated interactions, while all interventions universally improved willingness to initiate negotiations. However, regression analysis revealed a critical trade-off: \trc{} enhanced outcomes for structured negotiation approaches but sometimes undermined creative strategies, suggesting that incremental theoretical guidance conflicts with exploratory styles. Linguistically, \trc{}'s intentionally challenging yet accessible design—lower empathy but higher readability and theory integration—explains this pattern. The core finding: theory-driven coaching creates a preparation paradox where realistic challenge effectively reduces anxiety through structured exposure but may constrain confidence in creative contexts. This suggests that AI negotiation coaching influences psychological readiness most positively when theoretical scaffolding aligns with structured preparation needs, but may require design adaptations to support diverse negotiation approaches.

\begin{table*}[t]
\centering
\sffamily
\footnotesize
\caption{Comparison of lexico-semantic measures in the language of interactions with \trc{} and \cgpt{}. }
% Control vs. Trucey conditions. Control N = 66, Trucey N = 134, Degrees of Freedom = 198 for all tests.}
\label{tab:lexico_semantics}
\begin{tabular}{lrrrrrrrr}
% \hline
\textbf{Metric} & \textbf{\trc{}} & \textbf{\cgpt{}} & \textbf{Mean $\Delta$\%} & \textbf{Cohen's $d$} & \multicolumn{2}{c}{\textbf{t-test}} \\
\toprule
\rowcollight \multicolumn{7}{l}{\textit{Linguistic Structure}}\\
% User: Readability Grade & 8.74 & 9.33 & -6.69 & 0.07 & 0.50 & 0.62\\
% User: Flesch–Kincaid: Reading Ease & 72.77 & 70.74 & 2.79 & -0.09 & -0.61 & 0.54\\
Verbosity: AI & 135.68 & 363.45 & -167.88 & -4.14 & 27.55 & ***\\
Verbosity: User & 46.89 & 56.54 & -20.58 & -0.37 & 2.45 & **\\
\hdashline
Readability: AI & 80.00 & 68.46 & 14.43 & 2.17 & -14.42 & ***\\
Readability: User & 72.77 & 70.74 & 2.79 & 0.09 & -0.61 & \\
\hdashline
Repeatability: AI & 0.65 & 0.70 & -8.55 & -1.17 & 7.77 & ***\\
Repeatability: User & 0.43 & 0.38 & 11.30 & 0.42 & -2.76 & **\\
% User: Verbosity & 207.96 & 148.03 & 28.82 & -0.51 & -3.38 & 0.00\\
\hdashline
Complexity: AI & 3.91 & 4.02 & -2.84 & -0.67 & 4.45 & ***\\
Complexity: User & 3.98 & 3.94 & 0.95 & -0.10 & -0.65 & \\
% AI: Flesch-Kincaid Grade & 5.36 & 7.38 & -37.73 & 2.01 & 13.35 & 0.00\\
% assistant_verbosity & 849.16 & 1,286.98 & -51.56 & 1.15 & 7.62 & 0.00\\
\rowcollight \multicolumn{7}{l}{\textit{Linguistic Style}}\\
Categorical Dynamic Index (CDI): AI & 9.57 & 13.66 & -29.93 & -0.76 & -10.32 & ***\\
Categorical Dynamic Index (CDI): User & 14.52 & 15.07 & -3.65 & -0.04 & -0.59 & \\
\hdashline
Politeness: AI & 0.98 & 0.97 & 1.55 & 0.56 & -3.74 & ***\\
Politeness: User & 0.94 & 0.92 & 1.80 & 0.21 & -1.39 & \\
\hdashline
Formality: AI & 0.68 & 0.82 & -20.98 & -1.73 & 11.49 & ***\\
Formality: User & 0.79 & 0.81 & -2.43 & -0.12 & 0.81 & \\
\hdashline
Empathy: AI & 0.72 & 0.80 & -11.36 & -1.35 & 8.99 & ***\\
Empathy: User & 0.83 & 0.81 & 1.75 & 0.23 & -1.55 & \\
\hdashline
Persuasiveness: AI & 0.88 & 0.87 & 2.10 & 0.66 & -4.40 & ***\\
Persuasiveness: User & 0.88 & 0.87 & 1.13 & 0.20 & -1.33 & \\
\rowcollight \multicolumn{7}{l}{\textbf{Linguistic Adaptability}}\\
% \hdashline
% CDI Difference (AI - User) & -4.95 & -1.41 & 250.35 & -0.26 & -3.59 & ***\\
% \hdashline
Linguistic Style Accommodation & 0.57 & 0.60 & -5.16 & -0.08 & -1.10 & \\
\bottomrule
\end{tabular}
\Description[table]{This table compares the linguistic characteristics of interactions with the Trucey and ChatGPT AI systems. The analysis, which measures both AI and user language, focuses on metrics related to Linguistic Structure (such as verbosity, readability, and complexity) and Linguistic Style (including formality, empathy, and persuasiveness). The results show that Trucey's responses were significantly less verbose, more readable, and less complex than ChatGPT's. In terms of style, Trucey's language was also found to be more polite, less formal, more dynamic, and more persuasive, while being less empathetic than ChatGPT's.
}
\end{table*}

%% file: 4findings_rq2.tex
\section{RQ2: Participant Perspectives About (AI-Driven) Negotiation Coaching}

% \veda{I am trying to follow the structure in CarePilot as this is my first time writing and felt like I was benefiting from it's structure. I also know this has to be shorted but for now I have left it here for all of us to see the logic behind the argument as I cut down}

We qualitatively analyzed our semi-structured interviews data. The interviews were transcribed by an AI companion and manually reviewed by the researchers. 
First, we conducted an exploratory inductive coding to develop a codebook, and then conducted reflexive thematic analysis~\cite{braun2019reflecting}. The analysis of our semi-structured interview data allowed us to understand user experiences with different coaching approaches. Through iterative reading and comparison across transcript, three major themes with distinct sub themes.

This hierarchical analytical framework enabled the identification of participants' strategic orientations and the underlying rationale for their chosen approaches during AI coach interactions.
This analysis revealed how participants' experiences and interactions were influenced by psychological barriers (if any). 
We found three major themes: 1) Preference for informational autonomy, 2) Cognitive load management under preparation stress, and 3) Rehearsal acceptance divide.
We describe these themes and corresponding sub-themes in the following subsections. 
% \koustuv{needs to be updated!}
% 1) individual preparation paradigms, 2) tool integration, 3) workplace contextual constraints, and 4) emotional regulation mechanisms. 
% We conducted thematic analysis of semi-structured interviews with participants to understand how they interpreted and navigated the AI coach interventionns. 
% user experience landscape and how the interactions influenced the psychological barriers experienced by participants. The analysis reveals themes of individual preparation paradigms, tool integration, workplace contextual constraints, and emotional regulation mechanisms. 

% \subsection{Informational Autonomy}
\subsection{Preference for Informational Autonomy} \label{sec:autonomy_preference}
% Across all participants, the fundamental driver of preferences for a negotiation coach, was the need for \textbf{informational autonomy}---the ability to maintain control over when, how, and at what pace they consumed preparation guidance~\cite{}. 
% We found two distinct strategies that the participants adopted while navigating through the interventions:

Research on learner autonomy has demonstrated that satisfying user's autonomy needs enhances both satisfaction and learning effectiveness \cite{hu2017pathway}. Our analysis revealed that informational autonomy, the ability to maintain control over when, how and at what pace users consumed preparation guidance, emerged as a fundamental driver to a user's coaching tool preference. This superseded any technical features or interactive capabilities. 

Participants consistently emphasized their need to ``skim through everything'' and access ``bullet points'' for effective navigation. This autonomy manifested through two distinct strategies that participants used to maintain agency over their preparation processes, each reflecting relationships with technology-mediated learning. 

% However, participants achieved this autonomy through two distinct strategies that directly explain the observed tool performance patterns.

%\subsubsection{Maximum Autonomy Through Handbook.}
\subsubsection{Maximum Autonomy Through Comprehensive Control}

Six participants (P6, P7, P8, P10, P12, P15)  prioritized tools that provided immediate and unmediated access to expert knowledge. 
They valued the \hbk{}'s structural affordances for complete user control, viewing external mediation as a barrier rather than a benefit. Their preference reflected deeper needs for informational autonomy---direct possession of knowledge rather than a technology- dependent access.% through technology. 

Participants described wanting ``everything in front of you'' when comparing \hbk{} to the AI-driven interaction (\trc{} and \cgpt{}). P15's spatial metaphor captured this preference: comparing the \hbk{} to ``having a printed map versus Google Maps'' where comprehensive visibility of options trumped interactive guidance. This complete control allowed users to navigate preparation content according to their own cognitive rhythms and priorities.

The rejection of AI interaction requirements, system-driven engagement, and user-driven preparation. Participants characterized their approach through ``personality-based'' preferences, with some explicitly stating they ``can't endlessly chat with an AI assistant.'' Rather than engaging with interactive processes, these participants preferred to ``just read this and call it a day,'' viewing additional interactions as cognitive overhead rather than value-added guidance.

This pattern reflects what Fischer characterizes as a shift from \textit{passive consumer mindsets} toward informed participation~\cite{fischer1998beyond}. Rather than accepting AI-mediated guidance as passive recipients, participants sought direct access to comprehensive knowledge that enabled active, self-directed preparation. Their preference for informational autonomy represents informed participation, a need to take control of both content access and problem-solving strategies rather than following system-imposed workflows. 

For example, P12 showed interest in receiving the handbook, demonstrating an active stance toward knowledge acquisition, moving beyond consumption toward empowered engagement with preparation resources:

% P12's request to own the handbook file - 
 \begin{quote}
     \small
     ``I want this [negotiation Handbook] file. Do you mind sending it to me?'' --P12
 \end{quote}
\subsubsection{Strategic Autonomy Through Controlled Technology Integration}

Four participants (P11, P13, P14, and P15)  expressed an alternative autonomy strategy that extends~\citeauthor{fischer1998beyond}'s informed participation framework in a different direction~\cite{fischer1998beyond}. Rather than rejecting AI interventions entirely, they developed sophisticated methods to maintain agency while strategically leveraging current AI capabilities. These users demonstrated that empowered engagement with technology can be performed through controlled integration approaches that preserve user sovereignty over the preparation process.
% Participants 
They emphasized the importance of maintaining autonomy while interacting with AI systems, positioning them not as sources of passive guidance but as controlled tools for generating choices. This approach was particularly highlighted when users would like \cgpt{}, which gave them ``more control'' as it gave them ``multiple options.'' 
This strategy highlights how participants actively chose to express interest in partaking in systems that provided multiple pathways, allowing them to exercise informed choice among AI-generated alternatives. 

Furthermore, participants presented their thoughts on sophisticated hybrid workflows that allowed them to maintain autonomy over content selection and structural sequencing. P15 described a systematic approach: ``I'd start with the Handbook, and then practice with Trucey three or four times.'' 
This demonstrates how participants established foundational knowledge before engaging with interactive AI systems, maintaining control over timing and frequency. For instance, P13 outlined a strategic multi-system sequencing: 
\begin{quote}
    \small
    ``I would first like a framework with headings, then \cgpt{} when it provides more details, and then finally, the simulation part.'' --P13
\end{quote} 

Similarly, P14 emphasized targeted engagement as they would like  \textit{``to hone in to specific areas of concern.''}
These strategies demonstrate how participants maintained agency by strategically deploying different AI capabilities for specific preparation objectives rather than following system-designed workflows.

Together, these strategies demonstrate that information autonomy was one of the fundamental driver of coaching tool preference, creating two distinct but complementary approaches. Whether through direct knowledge ownership or controlled AI integration, participants consistently prioritized maintaining agency over their preparation process, showcasing a move from  passive consumption towards informed participation \cite{fischer1998beyond}.

While AI systems were positioned as flexible, controlled components, the handbook uniquely satisfied the dominant autonomy need for complete user control. This pattern validates prior research showing that satisfying autonomy needs enhances both satisfaction and learning effectiveness, extending these findings to workplace AI coaching contexts. The handbook's dominance in PEU (\autoref{tab:did}) observed earlier reflects the centrality of informational autonomy in structuring negotiation preparation, regardless of users' specific autonomy strategy.

\subsection{Cognitive Load Management Under Preparation Stress}
Research on fear and information seeking demonstrates that when individuals cannot avoid aversive events, they seek information to mitigate anticipated stress, with complete information reducing fear more effectively than partial information \cite{restrepo2023effect}. Cognitive load theory further indicates that anxiety and stress reduce working memory capacity, making information processing more challenging precisely when support is most needed  \cite{paas2003cognitive}.  

Our participants demonstrated acute sensitivity to information density and presentation format, particularly when experiencing negotiation-related anxiety in preparation.

\subsubsection{Stress-Sensitive Information Processing Requirements}
Four participants (P6, P8, P10, P13) explicitly connected information density to their emotional state and preparation effectiveness. They valued condensed, visually organized content that minimized cognitive processing burden during high-stress preparation periods. 
Participants expressed strong preferences for ``skimmable'' content and ``bullet points'' over verbose formats.  
P10 contrasted format effectiveness:

\begin{quote}
    \small
    ``\cgpt{} is hard to get info in that wall of text. Here, it's all points. You can skim effectively. The \hbk{} tells you the same thing with less fluff.'' --P10
\end{quote}

The stress-sensitive nature of negotiation preparation reveals deeper psychological mechanisms behind information preferences. Research shows that individuals facing unavoidable aversive events seek comprehensive information to reduce anxiety, with complete information proving more effective than partial information in mitigating fear responses~\cite{restrepo2023effect}. Additionally, P10 connected verbosity to emotional stating that if they're anxious, they wouldn't like to ``read such long messages.'' 
P13 expressed similar constraints, ``don't have time to read this much.'' 
These responses reflect an anxiety driven need for immediate, comprehensive information access rather than effortful information construction through AI dialogue, which requires higher information processing. 

This pattern suggests that cognitive constraints become more pronounced under preparation anxiety, with perceived time pressure compounding cognitive load challenges and making concise presentation crucial for tool acceptance during high-stress periods. 

\subsubsection{Information Architecture and Workflow Alignment Needs}

Two participants (P13, P14) demonstrated sensitivity to information sequencing and search optimization, revealing expectations about how guidance should align with natural preparation workflows rather than system-convenient delivery patterns.

P13 critiqued AI systems' temporal mismatch: "When I'm starting the conversation, the first message is also talking about following up [after negotiation] and be open to feedback. These things should come much later." This highlights the importance of information architecture that matches users' mental models of preparation sequences rather than system-convenient delivery orders.

Participants also valued tools that reduced "mental burden" and "effort" in locating relevant information. They preferred a minimized cognitive overhead in information seeking with a higher preference to comprehensive information tools, contrasting AI's sequential delivery with the handbook's efficient search capabilities.

These preferences challenge AI design approaches emphasizing scaffolding and incremental knowledge building. When participants expressed they ``feel like I don't have time to read this much'' or preferred to ``just read this and call it a day,'' they rejected interaction patterns that scaffolding AI requires. 

These findings reveal a fundamental mismatch between current AI design thinking and user preferences in high-stakes preparation contexts. Current approaches emphasize ``incremental construction of context and ideas through conversation'' and ``small information chunks'' to reduce cognitive overload~\cite{brickhouse1994plato}. 
This approach assumes gradual information building creates more ``cognitively manageable and emotionally sensitive dialogue.'' However, our participants directly contradicted these assumptions, seeking immediate access to comprehensive information rather than conversational layering. They preferred to ``just read this and call it a day'' rather than engage in sustained dialogue.

\subsection{The Rehearsal Acceptance Divide: Authenticity Versus Technical Capability}

The concept of AI-mediated rehearsal revealed fundamental philosophical differences about the nature of negotiation preparation and the role of simulation in skill development. Research on demonstrates that complex skills develop through deliberate practice with systemic variation and simulation based learning produces large positive effects on skill development~\cite{dekeyser2020skill,chernikova2020simulation}. Yet, our participants demonstrated a clear divide regarding whether AI systems could provide authentic enough practice experiences. This divide centered not on system quality but on deeper questions about whether AI-mediated practice could address the inherently human and unpredictable aspects of workplace negotiations.

% skill acquisition demonstrates that complex skills develop through repeated, deliberate practice with systematic variation~\cite{dekeyser2020skill}. Additionally, meta-analytic evidence shows that simulation based learning has known to produce large positive effect on skill development~\cite{chernikova2020simulation}. 

\subsubsection{Conditional Acceptance Through Technical Enhancement Pathways}

Five participants (P6, P9, P13, P14, P15) accepted rehearsal as conceptually valid but identified specific limitations that prevented current systems from reaching their potential. They consistently requested more  ``challenging scenarios'' and ``stricter pushbacks'' rather than systems that were ``too optimistic.'' , believing \trc{} was too ``agreeable''. P6 emphasized this need ``I want \trc{} to be like you cannot say no to me.'' 
This desire for realistic difficulty reflected their understanding that effective rehearsal should challenge and frustrate them, simulating the difficult dynamics rather than providing artificial optimism.

% wanted challenging, realistic scenarios that would prepare them for difficult dynamics rather than artificial optimism.
% Participants consistently requested more ``challenging scenarios'' and ``stricter pushbacks'' rather than systems that were ``too optimistic.'' They believed \trc{} was too ``agreeable'' and would like more adversarial preparation that would  push back aggressively. For instance, P6 emphasized this need for adversarial preparation, ``I want \trc{} to be like you cannot say no to me.''
% P6 emphasized this need for adversarial preparation: 
% \begin{quote}
%     ``I want \trc{} to be like you cannot say no to me.'' --P6
% \end{quote} 

These participants conceptualized rehearsal as requiring iterative practice with consistent personality modeling rather than single interactions. They wanted AI systems that would challenge and frustrate them, simulating the difficult dynamics they expected in real workplace negotiations. Furthermore, they framed rehearsal as requiring repeated practice and exposure to varied scenarios to build confidence and adaptability, equating to ``mock interviews.'' However, this was often contingent on the technical accuracy of the roleplay. 

The conditional acceptance was reinforced by our quantitative findings, showcasing how exposure through interaction with \trc{}, led to a reduction in fear. This aligns with research that varied stimulus exposure in controlled environments leads to fear reduction~\cite{rowe1998effects}. 
The observed anxiety reduction appeared to reflect a Dunning-Kruger correction, where initial overconfidence gave way to more accurate self-assessment after experiencing negotiation complexity~\cite{dunning2011dunning,he2023knowing}.
Rather than abandoning rehearsal, participants responded by seeking more challenging scenarios that matched their revised understanding of negotiation difficulty.

\subsubsection{Fundamental Rejection Despite Transfer Learning Evidence}

Six participants (P5, P7, P8, P10, P11, P12) explicitly rejected rehearsal regardless of potential benefits, perceiving AI-mediated practice as fundamentally inadequate for negotiation preparation. Prior work has shown that simulation effects do not guarantee transfer to real-world performance contexts~\cite{grierson2019simulation}, and this transfer uncertainty appeared to shape participants' evaluation of rehearsal value.

Their rejection centered on concerns that AI could not simulate the ``unpredictability'' and embodied psychological states inherent in workplace negotiations. Participants emphasized that AI failed to account for scenarios where people might ``panic, shut down, or freeze'' under pressure, or when someone might go into ``fight or flight mode'' during actual conversations.
P10 captured this core limitation:
\begin{quote}
\small
    ``It doesn't simulate scenarios like you might potentially panic or you might shut down, you might freeze. If you go into a fight or flight mode in the middle of the actual conversation, no AI can predict that.'' --P10
\end{quote}

Beyond physiological responses, participants questioned whether rehearsal could address the fundamental ``uncertainty of what will actually happen'' in real negotiations, arguing that even extensive rehearsal sessions represented only narrow subsets of possible workplace dynamics. Participants also experienced discomfort with AI interaction itself, trusting their own experiential knowledge over simulation and viewing rehearsal as introducing ``unnecessary complexity''. Some described feeling ``framed'' by systems that created bidirectional prompting relationships, with one noting that ``the simulation part pissed me off.''
% They argued that even extensive ``multiple rehearsal'' sessions represented only narrow subsets of possible workplace dynamics, failing to prepare them for the full range of negotiation outcomes.
% Apart from technical limitations, participants experienced fundamental discomfort with AI interaction itself. They trusted their own experiential knowledge as providing more reliable foundations than any simulation, believing they were "better judges" of how their managers and colleagues actually behave.
% Others viewed rehearsal as introducing ``unnecessary complexity'' that would require more ``time explaining to the system'' than conducting self-directed preparation. Some experienced deeper discomfort with the interaction format itself, describing feeling ``framed'' by systems that created bidirectional prompting relationships. This unease escalated to outright frustration, with one participant noting that ``the simulation part pissed me off.''
This rejection pattern reveals that for these participants, authentic preparation required human interaction and real-world unpredictability that could not be meaningfully reproduced through simulation, regardless of technical sophistication or demonstrated benefits such as fear reduction.

Individuals perceive theory driven AI negotiation coaching tools through the lens of informational autonomy and preparation context. Despite, \trc{}'s theoretical grounding, participants prioritized maintaining control over their process, often preferring access to immediate comprehensive information access over AI-mediated guidance. User acceptance of AI solutions hinged on autonomy preservation and cognitive load management under stress than on theoretical sophistication with a fundamental divide emerging around whether AI could authentically simulate workplace negotiation dynamics.

%% file: 5discussion_v2.tex
\section{Discussion}
The workplace is rapidly evolving with the adoption of AI to support professional communication, from drafting emails to rehearsing difficult conversations. Our work situates the AI-powered negotiation coach (\trc{}) within this ecosystem of AI-Mediated Communication (AIMC), extending its scope to negotiation rehearsal and psychological preparedness. 
While the popular assumption and hype emphasize AI's ability to enhance efficiency and personalization, our findings reveal a more nuanced picture: the benefits of AI in negotiation preparation are highly contingent on users' needs, emotional states, time constraints, and perceptions of authenticity. Contributing to HCI and CSCW debates on AI in socially complex, high-stakes interactions, we show that an AI's value is situated rather than universal. 
We discuss our implications through four lenses: reframing cognitive load theory for stress-sensitive contexts, examining authenticity and adaptation as boundaries of AI acceptance, considering methodological, social, workplace, and policy implications, and deriving design implications for autonomy, scaffolding, holistic preparation, and adaptive architectures.

\subsection{Reframing Cognitive Load Theory and AI's Situated Utility in Stress-Sensitive Contexts}

% Furthermore, o
Our findings highlight the utility of traditional, low-interaction learning modalities. Analogous to reading a textbook, a comprehensive \hbk{} often imposes intrinsically lower cognitive load for direct information acquisition~\cite{nguyen2022user}. 
This theoretical contribution reframes the optimal role of AI as being situated rather than universal: its value lies in easing extraneous cognitive factors in specific contexts, rather than always replacing direct content access.

Extending traditional cognitive load theory~\cite{sweller2011cognitive}, which primarily focuses on the germane load, our study contextualized a critical, underexplored dimension in human-AI interactions---the influence of stress-sensitive, anxiety-inducing contexts on user tolerance. Participants described how AI interaction requirements (e.g., query formulation, response parsing) imposed significant cognitive load, which was exacerbated by emotional states and time constraints. Our regression analyses (\autoref{sec: Regression Modeling}) support this observation: while structured AI guidance generally improved self-efficacy and empowerment, it can increase cognitive load in scenarios needing integrative or creative thinking, highlighting that stepwise support may conflict with users' need for flexible, open-ended reasoning under stress.
% Extending traditional cognitive load theory~\cite{sweller2011cognitive}, which primarily focuses on the germane load, our study contextualized and revealed a critical, underexplored dimension in human-AI interactions---the influence of stress-sensitive, anxiety-inducing contexts on user tolerance. % for AI interaction burden. 
% Participants described how AI interaction requirements (e.g., query formulation, response parsing) imposed significant cognitive load, which was exacerbated by emotional states and time constraints. \ziang{do we have quant evidence to support this?} 
This finding challenges the assumption that AI universally reduces effort or enhances efficiency in learning and preparation tools.

Consequently, our work urges researchers in human-AI collaborative systems to rethink traditional cognitive load frameworks to account for this dynamic interplay of emotional states and time constraints. In light of these findings, our work advocates for a new pedagogical lens that emphasizes a theoretical framework prioritizing cognitive load optimization by actively managing extraneous factors. This reorientation is crucial for both designing more effective AI systems and fostering a nuanced understanding of how cognitive resources are allocated and managed across high-stakes contexts, presenting a wide spectrum of user emotional and temporal conditions.

% Furthermore, our findings highlight the enduring utility of traditional, low-interaction learning modalities. Analogous to reading a textbook, a comprehensive \hbk{} often imposes intrinsically lower cognitive load for direct information acquisition~\cite{nguyen2022user}. 
% This theoretical contribution reframes the optimal role of AI as being situated rather than universal: its value lies in easing extraneous cognitive factors in specific contexts, rather than always replacing direct content access. 
% \ziang{I love this paragraph but I am not sure if it should be the first paragraph in discussion}

\subsection{AI's Situated Role: Authenticity, Acceptance, and Socio-Technical Adaptation}
% \ziang{I think we can cite papers that talk about the fidelity of simulation. A few examples, \cite{zou2025can,cheng2023compost}. Cheng's paper argues the LLM's simulation is lack of diversity and my paper argues the gap between eval and real interaction. From a technical perspective, there are limitations of using LLM to run simulation which could be the source lack of authenticity}

Our findings reveal a critical theoretical boundary in AI acceptance, contributing to theories on user acceptance, particularly in simulation fidelity~\cite{grierson2019simulation,cheng2023compost}. 
Prior work shows that the effects of simulation do not necessarily transfer to real-world contexts~\cite{grierson2019simulation}, and recent research highlights the limitations of LLM-based simulations---such as lack of diversity~\cite{cheng2023compost} and gaps between evaluation settings and real interactions~\cite{zou2025can}. Participants in our study echoed similar concerns, perceiving AI-mediated rehearsal as inadequate not only for technical reasons but also due to psychological barriers tied to the authenticity of practice experiences. This underscores the need to investigate how authenticity concerns manifest across preparation contexts and whether hybrid approaches (e.g., combining AI with peer rehearsal) might mitigate these limitations.
% Participants perceived AI simulation inadequacy not as a technical challenge but also as a psychological barrier tied to the authenticity of practical experiences. This underscores the need to study how authenticity concerns manifest across different preparation contexts and whether hybrid approaches (e.g., combining AI with peer rehearsal) can address these barriers.

Yet, despite these authenticity-related limitations, AI coaching provides unique preparation advantages. Our findings in Section(~\ref{sec: Regression Modeling}) demonstrate measurable benefits, with tools such as \trc{} significantly improving user's confidence, reducing fear, and increasing willingness to initiate a negotiation. These were particularly strong in strategic contexts, where AI tools can help break down problems and allow users to perceive information dynamically. 
Furthermore, AI also offers consistent availability for practice, especially in sensitive workplace scenarios that either cannot easily be shared with colleagues or friends or in which they may be unwilling or uncomfortable to engage.
Additionally, AI can reliably provide challenging, adversarial scenarios without social consequences, something human practice partners often avoid to preserve relationships. Finally, AI addresses scalability and accessibility barriers, providing structured preparation support to workers who lack access to professional coaching or willing practice partners.

Taken together, our work calls for rethinking how AI adaptation mechanisms can be designed 
% prompting strategies must be designed 
for socially complex, high stakes interactions. While current LLM prompting techniques optimize for structured reasoning, workplace negotiations demand sensitivity to power asymmetry, relational dynamics, and emotional nuances. By embedding these dimensions directly interaction design enables adaptation to function as a
% into prompts, prompting becomes a 
mechanism for operationalizing social theories within AI-mediated interactions. %\ziang{instead of prompting strategies, I think adaptation might be a better term}
Therefore, our work argues for theorized adaptation practices---moving beyond technical tuning toward encoding social science insights into input-output structuring---so that AI can better scaffold negotiation readiness in authentic, context-sensitive ways.
% as a design practice, moving beyond technical tuning toward encoding social science insights at the level of input–output structuring.

\subsection{Methodological, Policy, and Workplace Implications}

From a methodological perspective, our work provides preliminary insights into how prompt engineering can serve as a lightweight, cost-effective alternative to model fine-tuning and retraining. This is particularly important given the resource constraints many organizations and researchers face in customizing large models for specific social use cases. By embedding domain knowledge, social theory, and user context directly into prompts, we show how meaningful interactional shifts can be achieved without modifying the underlying model. 
This approach not only enhances customization but also supports rapid prototyping and deployment of socially attuned AI tools across diverse workplace settings. 
Methodologically, this points toward a broader agenda in CSCW and HCI research: treating prompts as socio-technical scaffolds that mediate interaction, rather than as mere technical levers for content generation.

Further, our study offers a reusable and adaptable approach for AI-mediated rehearsal in professional contexts---such as preparing for difficult conversations with managers or navigating conflict in team settings. 
Extending beyond negotiation, this approach can also benefit domains such as career coaching, leadership training, or mental health support, where users must manage stigma, relational asymmetries, and emotionally fraught interactions. At the same time, social risks remain. AI-generated outputs are known to reproduce patterns of linguistic and social bias from their training data~\cite{bender2021dangers,hohenstein2023artificial}. In negotiation contexts, this could skew the way authority, assertiveness, or politeness are modeled, shaping users' sense of agency or reinforcing harmful communicative patterns. Moreover, such technologies may disproportionately benefit individuals with high digital fluency or culturally dominant communication styles, while others---such as non-native speakers or underrepresented groups—may struggle to align their intentions with biased outputs or develop over-reliance on AI-generated suggestions.

From a workplace perspective, these dynamics highlight both opportunities and challenges for AI-mediated communication (AIMC)~\cite{corvite2022data,das2023algorithmic,kawakami2023wellbeing,kaur2022didn,dasswain2025ai}. 
Tools like \trc{} can scaffold negotiation preparation, boost confidence, and reduce fear, contributing to more equitable participation in high-stakes interactions. 
Yet they also raise concerns around bias, trust, accessibility, and cultural variance. 
Organizational hierarchies and cultural norms shape what counts as ``appropriate'' negotiation---whether assertiveness, hierarchy, or collaboration---and AI systems not calibrated to these contexts risk privileging dominant strategies while marginalizing others~\cite{das2023algorithmic,das2024teacher}. 
Importantly, AI negotiation tools should not be seen as productivity enhancers for employers, but as user-facing supports for empowerment, inclusion, and resilience. 
Their value lies in enabling workers to prepare more confidently and equitably, while organizational policies and infrastructures can play a supporting role in ensuring such tools are accessible, trustworthy, and responsibly integrated into professional development ecosystems.

Taken together, these implications suggest that the development of AI coaching tools for workplace communication must proceed with attention to methodological rigor, social equity, and institutional accountability. Methodologically, lightweight but theory-informed prompt engineering offers a promising pathway for rapid, context-sensitive customization. Socially, systems must be designed to mitigate bias, support marginalized users, and accommodate diverse cultural norms of communication. At the workplace level, AI negotiation tools should be understood not merely as productivity aids, but as user-facing supports for empowerment, inclusion, and resilience. Their value lies in enabling workers to prepare more confidently and equitably for high-stakes interactions, while organizational policies and training infrastructures can play a supporting role in ensuring such tools are accessible, trustworthy, and responsibly integrated into professional development ecosystems.

\subsection{Design Implications}

\subsubsection{Designing for Autonomy Needs }
% \ziang{not sure if Maximum is the best term, should we just say Autonomy Needs?}
Participants operated under specific contextual pressures that revealed fundamental mismatches between conversational AI and workplace preparation needs (\autoref{sec:autonomy_preference}). 
When preparing for negotiations with supervisors, managing performance anxiety, and working within constrained windows, participants preferred comprehensive, readily accessible materials over conversational engagement. They wanted direct informational access that could be stored, reused, and revisited, rather than fragmented back-and-forth dialogue. 
This underscores a critical need for AI coaching tools to provide efficient, contextualized knowledge delivery tailored for high-stakes negotiation preparation.

Literature on repository-based knowledge systems has shown how information retrieval and content curation can support autonomy and re-usability~\cite{hearst2009search}. Such systems could prompt users for explicit situational parameters (e.g., power dynamics, relationship type, stakes level) rather than relying solely on automatic inference, thereby increasing reliability and transparency. Based on these user inputs, AI can generate multiple options organized by theoretical frameworks---for example, integrative methods for collaborative contexts, distributive strategies for competitive situations, and principled approaches for complex dynamics~\cite{fisher2011getting}. 
This preserves user autonomy while leveraging AI's ability to curate and organize expert knowledge. Delivering this content in exportable formats (e.g., documents, slide decks) further supports offline access and long-term retention, which participants highlighted as essential for workplace preparation.

\subsubsection{Context-Sensitive Scaffolding for Cognitive Load Management}

Participants demonstrated stress-sensitive information processing patterns, highlighting a need for technology to adapt to cognitive load and time constraints. For example, a user with ample time may engage more deeply, while those facing immediate deadlines may express frustration with cognitively demanding interactions. 

To address these stress-sensitive needs and decrease cognitive load, our work inspires designing innovative AI coaching architecture featuring dynamic and context-sensitive scaffolding. This aligns with cognitive load theory~\cite{sweller2011cognitive} and research on guided instruction~\cite{kirschner2006unguided}, which show that structuring tasks reduces overload. 
Such tools can additionally incorporate flexibility modes, so that the user also has options to toggle what kind of information gathering they are looking for at a particular time. 
A binary scaffolding mode can be offered: streamlined versus detailed. The streamlined mode would deliver concise outlines of negotiation components and prioritized action steps, minimizing back-and-forth interaction for users already experiencing overload. The detailed mode would provide deeper theoretical grounding, case examples, and strategic frameworks for those with time to engage. Clear demarcation of preparation stages further reduces complexity, breaking down tasks into manageable units and surfacing only the most relevant information at each step. Similar adaptive scaffolding approaches have been explored in conversational AI for customer service~\cite{xu2017new}, suggesting feasibility in negotiation coaching contexts.% 
\subsubsection{Integrated Stages for Holistic Preparation and Psychological Resilience}

Building on capabilities such as strategic advice and AI-powered role-based simulation showcased by \trc{}, future technologies can integrate distinct stages for holistic preparation---addressing both tactical and psychological sides of negotiation. 
Beyond generating strategies, such tools can incorporate structured reflection sessions after practice interactions. 
These reflections can aim to empower users to process emotional responses to preparation scenarios, pinpoint specific moments where confidence wavered, and develop targeted strategies for subsequent preparation cycles.

By serving as a continuous repository of learning and self-reflection, AI coaching tools can transcend beyond short-term and immediate support tools to become longer-term skill-building assistants throughout an individual's professional journey. 
This dual focus on tactical mastery and resilience aligns with evidence that negotiation training has sustained effects only when reinforced longitudinally~\cite{zerres2013does}.
Such technologies can uniquely bridge unilateral self-preparation (repository and reflection) with bilateral interaction (simulation and feedback), optimizing for long-term learning outcomes. Integrating wellbeing-support modules, such as stress regulation strategies could further enhance psychological resilience, connecting to literature on emotion regulation in high-stakes work contexts~\cite{grandey2000emotional,dasswain2025ai}.

\subsubsection{Adaptive Information Architecture for User Autonomy and Learning Strategies}

Multiple participants emphasized a need for integrating multiple AI tools into a personalized preparation workflow. This motivates a design consideration for AI coaching systems to provide an adaptive information architecture.
Rather than prescribing a single interaction mode, AI tools can function more as intelligent, adaptable platforms where users can construct their own information pipelines---and such a design will also respect and empower users' autonomy in defining their learning strategies.

The above architecture prioritizes user agency by allowing explicit sequencing and combination of system components--- comprehensive handbooks, simulations, strategic guidance, theoretical framings, or reflection modules.
Research on learning personalization in intelligent tutoring systems supports this approach, showing that explicit control over scaffolding fosters greater motivation and engagement~\cite{aleven2003help}. 
For users preferring interactive engagement, the system could provide dynamic recommendations informed by topic modeling (e.g., urgent need for information) or tonal analysis (e.g., desire for empathy and reassurance). In this way, the system becomes responsive to both situational context and user state, ensuring content delivery aligns with immediate needs while preserving overall autonomy. Integrating with external ecosystems (e.g., calendars, email, or document repositories) would further embed coaching workflows into users' daily ~\cite{cranshaw2017calendar}.

%% file: 6limitations.tex
\subsection{Limitations and Future Directions}

Our work has limitations, which also suggest interesting directions for future research. We did not place participants in an actual negotiation experience. Although our approach provided controlled insights into user perceptions and preparatory behaviors, future studies could create AI simulations of managers or supervisors with reward and resistance strategies to approximate real negotiations. 
Such settings would allow for richer evaluation of how AI coaching holds up in authentic, multi-turn, and high-stakes exchanges.

Again, participants may not have had an immediate need to negotiate at the time of the study. Although this allowed us to capture baseline perceptions of usefulness, it also limits ecological validity. Future research should evaluate AI coaching with participants facing real, imminent negotiation demands, where motivation, stakes, and outcomes are directly relevant to their workplace contexts. As with many studies in this space, our experimental study relied on a crowd sample (from Prolific) which constrains the generalizability of findings, as crowd participants may differ from professionals actively engaged in workplace negotiations, particularly in organizational role, cultural background, and lived experiences of power asymmetry. 
To complement this, we conducted a smaller qualitative study with 15 participants to gain deeper insights into how individuals navigated the system. However, because recruitment occurred through our social media and our LinkedIn networks, this pool was disproportionately composed of individuals from engineering and technology backgrounds, further limiting diversity of perspectives. Extending this line of work with field studies in organizational contexts would provide stronger ecological grounding.

Finally, while our findings suggest promising directions for prompt engineering as socio-technical scaffolding, the effectiveness of this approach in more ambiguous, multi-turn real-world conversations remains an open question. Future work should extend evaluation through live user studies across diverse workplace negotiations, coupled with expert assessments of negotiation quality and outcomes. This line of research could also explore cross-cultural variation in role expectations, tone perceptions, and power dynamics, as well as integrate prompt scaffolding into collaborative workplace tools. More broadly, future studies should theorize how conversational AI can act not only as coaches but also as mediators in other sensitive and asymmetric social interactions, such as performance reviews or conflict resolution.

% The investigation was exploratory and scenario specific. \usc{}'s performance was evaluated only on a single controlled simulation and depended on well-structured user inputs. As a result, its effectiveness in more ambiguous, multi-turn real world conversations, remains an open question. 
% Future work shall extend this evaluation through live user studies across diverse workplace negotiations. Furthermore, these conversations will have to be examined by experts to evaluate the success and relevance of these results to users. 
% Our work inspires future research to expand on a similar approach by also exploring cross-cultural variation in role expectations and tone perceptions, as well as in integrating prompt scaffolding into collaborative tools. Finally, this line of work can also look into further theorizing how conversational AI can act as mediators in other sensitive and asymmetric social interactions.

% \koustuv{We did not put users into an actual negotiation experience. Future studies can create Ai simulations of managers with reward strategies to simulate such an experience and evaluate the effectiveness.}

% \koustuv{Participants voiced that even thouh they went through this exercise, they were actually not in a ``need'' to have a negotiation in the immediate future. }

% \koustuv{Limitations of working with a crowd-sample. }

%% file: 7conclusion.tex
\section{Conclusion}

This study studied AI's role in workplace negotiation preparation by empirically comparing a theory-driven AI coach (\trc{}), a general-purpose AI (\cgpt{}), and a traditional handbook (\hbk{}). Across $N$=267 participants and $N$=15 semi-structured interviews, we found that \trc{} significantly reduced negotiation-related fear, yet the handbook consistently outperformed both AIs in usability and psychological empowerment by providing comprehensive, reviewable content that bolstered confidence. 
Interviews further revealed a trade-off: participants valued AI's availability and rehearsal opportunities; however, its fragmented, verbose guidance often imposed additional cognitive burden, leaving them uncertain or overwhelmed. 
We discussed how our findings challenge assumptions of AI superiority and point toward hybrid designs that combine the structured clarity of handbooks with the adaptive rehearsal capabilities of AI. 
More broadly, our work reframes AI's role as situated and contingent, most valuable when systems reduce extraneous cognitive load, respect users’ psychological needs, and scaffold autonomy in high-stakes workplace interactions.